\documentclass[11pt,letterpaper]{article}
\usepackage[left=1 in, top=1 in, right=1 in, bottom=1 in]{geometry}

\usepackage{amsmath,amssymb}
\usepackage{amsfonts}
\usepackage{float}
\usepackage{changepage}

\usepackage[utf8x]{inputenc}

\usepackage{textcomp,marvosym}

\usepackage{cite}

\usepackage{nameref,hyperref}

\usepackage{array}

\newcolumntype{+}{!{\vrule width 2pt}}

\makeatletter
\renewcommand{\@biblabel}[1]{\quad#1.}
\makeatother

\date{}

\usepackage{bm}

\usepackage{enumitem}
\setlist[itemize]{noitemsep,topsep=0pt}
\usepackage{lineno}
\nolinenumbers
\parskip = \baselineskip

\usepackage{color}
\usepackage{amsfonts}
\usepackage{multicol}
\usepackage{fancyhdr}
\usepackage{graphicx}
\usepackage[numbers,sort&compress]{natbib}
\usepackage[aboveskip=1pt,labelfont=bf,labelsep=period,justification=raggedright,singlelinecheck=off]{caption}

\begin{document}

\begin{flushleft}
{\Large
\textbf\newline{Interactions between species introduce spurious associations in microbiome studies} }
\newline
\\
Rajita Menon\textsuperscript{1},
Vivek Ramanan\textsuperscript{2,4},
Kirill S. Korolev\textsuperscript{1,3*},
\\
\bigskip
\textbf{1} Department of Physics, Boston University, Boston, Massachusetts, USA
\\
\textbf{2} BRITE Bioinformatics REU Program, Boston University, Boston, Massachusetts, USA
\\
\textbf{3} Graduate Program in Bioinformatics,  Boston University, Boston, Massachusetts, USA
\\
\textbf{4} Department of Biology and Computer Science, Swarthmore College, Swarthmore, Pennsylvania, USA
\\
\bigskip

* korolev@bu.edu

\end{flushleft}

\section*{Abstract}
Microbiota contribute to many dimensions of host phenotype, including disease. To link specific microbes to specific phenotypes, microbiome-wide association studies compare microbial abundances between two groups of samples. Abundance differences, however, reflect not only direct associations with the phenotype, but also indirect effects due to microbial interactions. We found that microbial interactions could easily generate a large number of spurious associations that provide no mechanistic insight. Using techniques from statistical physics, we developed a method to remove indirect associations and applied it to the largest dataset on pediatric inflammatory bowel disease. Our method corrected the inflation of p-values in standard association tests and showed that only a small subset of associations is directly linked to the disease. Direct associations had a much higher accuracy in separating cases from controls and pointed to immunomodulation, butyrate production, and the brain-gut axis as important factors in the inflammatory bowel disease.

\section*{Introduction}
Microbes are essential to any ecosystem be it the ocean or the human gut. The sheer impact of microbial processes has however been underappreciated until the advent of culture-independent methods to assess entire communities~\textit{in situ}. Metagenomics and 16S rRNA sequencing identified significant differences in microbiota among hosts, and experimental manipulations established that microbes could dramatically alter host phenotype~\cite{hmp, yatsunenko:geography, cho:hmp_review, ding:enterotypes, gilbert:emp, bakken:c_dif, suez:glucose, jumpertz:energy}. Indeed, anxiety, obesity, colitis, and other phenotypes can be transmitted between hosts simply by transplanting their intestinal flora~\cite{turnbaugh:obesity_microbiome, messaoudi:anxiety_probiotic, cryan:behavior_microbiome, palm:antibody_coating, sampson:pd_microbiome}. 

New tools and greater awareness of microbiota triggered a wave of association studies between microbiomes and host phenotypes. Microbiome wide association studies~(MWAS) have been carried out for diabetes, arthritis, cancer, autism and many other disorders~\cite{mwas:diabetes, kostic:diabetes_microbiome, giongo:diabetes_microbiome, brusca:arthritis_microbiome, taneja:arthritis_microbiome, williams:sutterella_autism, wang:sutterella_autism, gevers:risk, saudi:fungi, metagenome:mwas}. MWAS clearly established that each disease is associated with a distinct state of intestinal dysbiosis, but they often produced conflicting results and identified a very large number of associations both within and across studies~\cite{williams:sutterella_autism, son:sutterella_autism, gevers:risk, wang:risk, de_cruz:ibd_review, mwas:diabetes, metagenome:mwas}. For example, a recent study on inflammatory bowel disease~(IBD) reported close to~100 taxa associated with IBD~\cite{wang:risk}, a number that is fairly typical~\cite{mwas:diabetes}. Such long lists of associations defy simple interpretation and complicate mechanistic follow-up studies because one needs to examine the role of almost every species in the microbiota. In fact, one can argue that MWAS are most useful when they can identify a small network of taxa driving the disease.

Although extensive dysbiosis might reflect the multifactorial nature of the disease, it is also possible that MWAS detect spurious associations because their statistical methods fail to account for some important aspects of microbiome dynamics. One such aspect is the pervasive nature of microbial interactions: species compete for similar resources, rely on cross-feeding for survival, and even produce their own antibiotics~\cite{coyte:stability, foster:cooperation_gut, flint:interactions, bashan:universality, faust:microbiome_interactions, magnusdottir:agora, chu:gut_antibiotics, riley:warfare, czaran:warfare, dethlefsen:assembly_review, mackie:interactions_gut}. Hence, microbial abundances must be correlated with each other, and even a simple change in host phenotype could manifest as collective responses by the microbiota. Traditional MWAS, however, completely neglect this possibility because they treat each species as an independent manifestation of host phenotype. As a result, MWAS cannot distinguish taxa directly linked to disease from taxa that are affected only through their interactions with other species.

The main conclusion of this paper is that realistic microbial interactions produce a large number of spurious associations between particular members of the microbiome and phenotypes. Many of these indirect associations can be removed by a simple procedure based on maximum entropy models from statistical physics~\cite{sander:maxent_review, bialek:biophysics}. We dubbed this approach Direct Association Analysis, or DAA for short.

When applied to the largest MWAS on IBD, DAA shows that many of the previously reported associations could be explained by interspecific interactions rather than the disease. At the genus and species level, the direct associations include only~\textit{Roseburia}, \textit{Faecalibacterium prausnitzii}, \textit{Bifidobacterium adolescentis}, \textit{Blautia producta}, \textit{Turicibacter}, \textit{Oscillospira}, \textit{Eubacterium dolichum}, \textit{Aggregatibacter segnis}, and~\textit{Sutterella}. Some of these associations are well-known~\cite{morgan:roseburia_down, machiels:butyrate_down, morgan:ibd_metagenomics, travis:roseburia_genome, forbes:gut_inflammatory_diseases, joossens:prausnitzii_low, sokol:prausnitzii_low, blautia:reduced_butyrate}, while others have received little attention in IBD research. The phenotypes of the taxa directly linked to disease suggest that immunomodulation, butyrate production, and the brain-gut interactions play an important role in the etiology of IBD. 

Compared to traditional MWAS, DAA corrected the inflation of p-values responsible for the large number of spurious associations and identified taxa most informative of the diagnosis. We found that directly associated taxa are much better at discriminating between cases and controls than an equally-sized subset of indirect associations. In fact, direct associations have the same potential to discriminate between health and disease as the entire set of almost a hundred associations detected by conventional methods.
\section*{Results} 
\begin{figure}[H]
\begin{center}
\includegraphics[width = 6in]{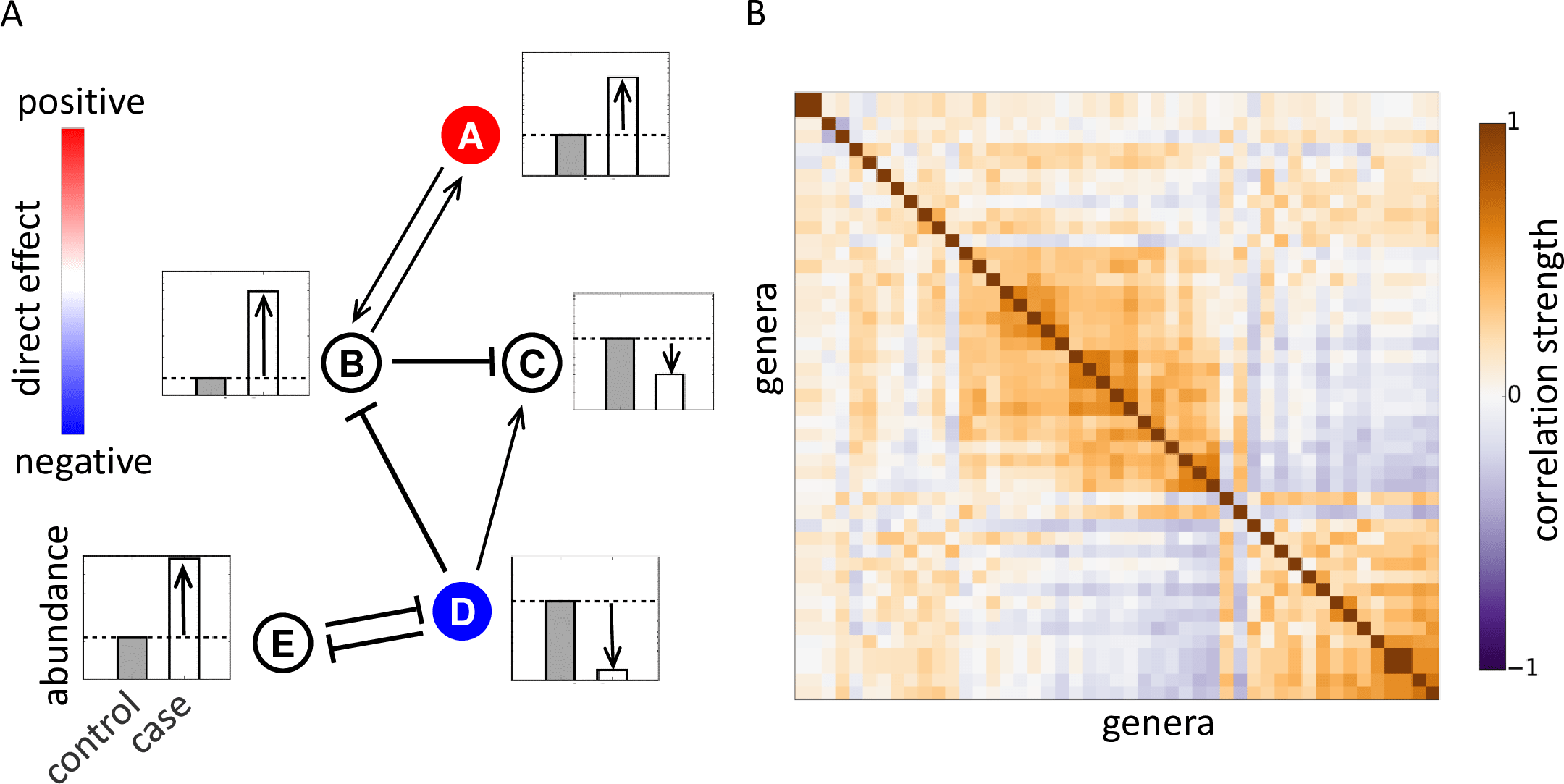}
\vspace{.25in}
\caption{{\bf Microbial interactions generate spurious associations.} 
\textbf{(A)}~A hypothetical interaction network of five species together with their dynamics in disease. Only two species~(shown in color) are directly linked to host phenotype. These directly-linked species inhibit or promote the growth of the other members of the community (shown with arrows). As a result, all five species have different abundances between case and control groups. \textbf{(B)} Microbial interactions are visualized via a hierarchically-clustered correlation matrix computed from the data in Ref.~\cite{gevers:risk}. We used Pearson's correlation coefficient between log-transformed abundances to quantify the strength of co-occurrence for each genus pair. Dark regions reflect strong interspecific interactions that could potentially generate spurious associations. See Supplementary Information for the list of 47~most prevalent genera included in the plot.}
\label{fig:Figure_1}
\end{center}
\end{figure}

Traditional MWAS detect species with significantly different abundances between case and control groups. Some changes in the abundances are directly associated with the disease while others are due to microbial interactions. The emergence of indirect changes in abundance is illustrated in Fig.~ \ref{fig:Figure_1}A for a hypothetical network of five species. Only two species~A and~D are directly linked to the disease. However, strong interactions make the abundances of all five species differ between control and disease groups. For example, the mutualistic interaction between~A and~B helps~B grow to a higher density following the increase in the abundance of~A. The expansion of~B in turn inhibits the growth of~C and reduces its abundance in disease. Strong mutualistic, competitive, commensal, and parasitic interactions have been demonstrated in microbiota~\cite{coyte:stability, foster:cooperation_gut, flint:interactions, bashan:universality, faust:microbiome_interactions, magnusdottir:agora, chu:gut_antibiotics, riley:warfare, czaran:warfare, dethlefsen:assembly_review, mackie:interactions_gut}, and Fig.~\ref{fig:Figure_1}B shows that almost every species present in the human gut participates in a strong interaction. Thus, the propagation of abundance changes from directly-linked to other species could pose a significant challenge for MWAS. To test this hypothesis, we turned to a minimal mathematical model of microbiota composition.
\vspace{.25in}
\vspace{-.5in}
\subsection*{Maximum entropy model of microbiota composition} A quantitative description of interspecific interactions and their effect on MWAS requires a statistical model of host-associated microbial communities. Ideally, such a model would describe the probability to observe any microbial composition, but the amount of data even in large studies is only sufficient to determine the means and covariances of microbial abundances. This situation is common in the analysis of biological data and has been successfully managed with the use of maximum entropy distributions~\cite{sander:maxent_review}. These distributions are chosen to be as random as possible under the constraints imposed by the first and second moments. Maximum entropy models introduce the least amount of bias and reflect the tendency of natural systems to maximize their entropy~\cite{plischke:statistical_physics, harte:maxent}. In other contexts, these models have successfully described the dynamics of neurons, forests, flocks, and even predicted protein structure and function~\cite{bialek:maxent_neurons, volkov:interactions, mora:maxent_flocks, morcos:direct, chakraborty:hiv_sectors}. In the context of microbiomes, a recent work derived a maximum entropy distribution for microbial abundances using the principle of maximum diversity~\cite{fisher:habitat_fluctuations}.

We show in the Supplementary Information that the maximum entropy distribution of microbial abundances~$P(\{l_{i}\})$ takes the following form

\begin{equation}
P(\{l_{i}\}) = \frac{1}{Z} e^{\sum_{i}h_{i}l_{i} + \frac{1}{2}\sum_{ij}J_{ij}l_{i}l_{j}}
\label{eq:maxent}
\end{equation}

\noindent where~$l_{i}$ is the log-transformed abundance of species~$i$, $h_i$ represents the direct effect of the host phenotype on species~$i$, and~$J_{ij}$ describes the interaction between species~$i$ and~$j$; the factor of~$1/Z$ is the normalization constant. The log-transformation of relative abundances alleviates two common difficulties with the analysis of the microbiome data. The first difficulty is the large subject-to-subject variation, which is much better captured by a log-normal rather than a Gaussian distribution; see \~Fig. \ref{fig:lognormal}, and Ref.~\cite{wang:risk}. The second difficulty arises from the fact that the relative abundances must add up to one. This constraint is commonly known as the compositional bias because it leads to artifacts in the statistical analysis~\cite{friedman:sparcc, aitchison:original, pawlowsky:book}. The log-transformation is an essential first step in most methods that account for the compositional bias including the widely advocated log-ratio transformation~\cite{friedman:sparcc, aitchison:original, pawlowsky:book, egozcue:isometric_scaling}, which includes additional steps that are not relevant in the context of Eq.~(\ref{eq:maxent}). In the Supplementary Information, we generalize Eq.~(\ref{eq:maxent}) to account for the constraint imposed by data normalization and show that our conclusions are not affected by the compositional bias.

The key prediction of Eq.~(\ref{eq:maxent}), see Supplementary Information, is that~$h$ and mean microbial abundances~$m_i=\langle l_i \rangle$ are related by~$m= J^{-1}h$. Because of interspecific interactions,~$J$ is not diagonal, and, therefore, a change in one component of~$h$ affects the abundances of many species. We show below that this nontrivial cause-effect relationship gives rise to spurious associations in both synthetic and real microbiome data. 
\vspace{-.25in}
\subsection*{Testing for spurious associations in synthetic data}
\begin{figure}[H]
\begin{center}
\includegraphics[width = 6.5in]{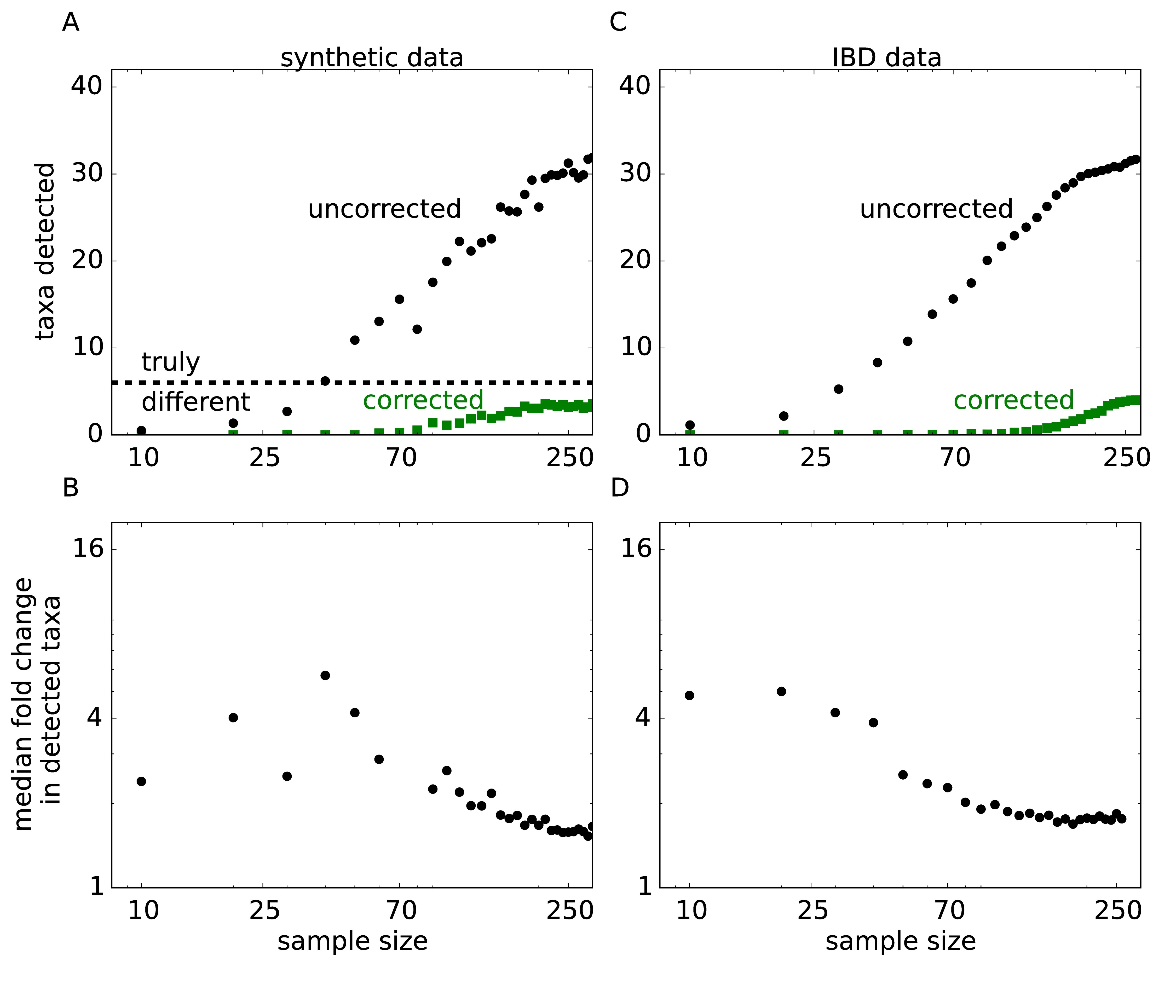}
	\caption{{\bf Signatures of indirect associations in synthetic and IBD data sets.} 
The synthetic data set was generated to match the statistical properties of the IBD data set from Ref.~\cite{gevers:risk}, but with a predefined number of 6 directly associated taxa (See Supplementary Information). \textbf{(A)}~In synthetic data, DAA identifies no spurious association and detects 4 out of 6 directly associated genera. All 6 genera and no false positives are detected when the sample size is increased further~(~Fig. \ref{fig:Figure2_full}). In sharp contrast, a large number of spurious associations is observed for metrics that rely on changes in abundance between cases and controls and do not correct for microbial interactions. The number of false positives grows rapidly with statistical power until all taxa are reported as significantly associated with the disease. \textbf{(B)} All spurious associations show substantial differences between cases and controls and, therefore, cannot be discarded based on their effect sizes. To quantify the effect size, we estimated the magnitude of the fold change for each genus. Specifically, we first computed the difference in the mean log-abundance between cases and controls and then exponentiated the absolute value of this difference. The plot shows how the median effect size for significantly associated genera depends on the sample size. Larger samples sizes result in much higher number of associations, but only a small drop in the typical effect size. (\textbf{C}) and~(\textbf{D}) are the same as~(A) and~(B), but for the IBD data set. The results are consistent between the two data sets suggesting that most associations detected by traditional MWAS are spurious. The complete list of indirect associations inferred from the IBD data set is shown in the Supplementary Information, and the results for different synthetic data sets are shown in ~Fig. \ref{fig:alternative_synthetic_effects1}.}
	\label{fig:Figure_2}
\end{center}
\end{figure}

\vspace{-.25in}
We obtained realistic model parameters from one of the largest case-control studies previously reported in Ref.~\cite{gevers:risk}. The samples were obtained from mucosal biopsies of 275 newly diagnosed, treatment-naive children with Crohn's disease (a subtype of IBD) and 189 matched controls. Microbiota composition was determined by 16S rRNA sequencing with about 30,000 reads per sample. From this data, we inferred the interaction matrix~$J$ and the typical changes in microbial abundances associated with the disease for 47 most prevalent genera (Methods and Supplementary Information). Even though the number of data points significantly exceeds the number of free parameters in the model, overfitting could still be a potential concern. However, overfitting is unlikely to affect our main conclusions because they depend only on the overall statistical properties of~$J$ rather than on the precise knowledge of every interaction. In fact, none of our results changed when we analyzed only about half of the data set~(Fig.~\ref{fig:Figure_2} and ~Fig. \ref{fig:dh_subsampled}). To improve the quality and robustness of the inference procedure, we also used the spectral decomposition of~$J$ to remove any interaction patterns that were not strongly supported by the data; see Methods and Supplementary Information for further details.

To determine the effect of microbial interactions on conventional MWAS analysis, we generated synthetic data with a known number of direct associations. The data for the control group was used without modification from Ref.~\cite{gevers:risk}. The disease group was generated using Eq.~(\ref{eq:maxent}) with the same values of~$h$ and~$J$ as in the control group, except we modified the values of~$h$ for~6 
representative genera~(see Supplementary Information). We also generated two other synthetic data sets with smaller and larger effect sizes. The results for all three data sets were very similar~(Supplementary Information).

The synthetic data was further subsampled to several sample sizes in order to simulate variation in statistical power between different studies. For an ideal method, the number of detected associations should increase with the cohort size, but eventually saturate once all~6 directly associated genera are discovered. In contrast to this expectation, the number of associations detected by the conventional approach increased rapidly with the sample size until almost all genera were found to be statistically associated with the disease in our synthetic data. At this point, traditional MWAS completely lost the power to identify the link between the phenotype and microbiota. Unbounded growth in the number of detections was also observed for the real data~(Fig.~\ref{fig:Figure_2}C) suggesting that many previously reported associations between microbiota and IBD could be indirect.

Are spurious associations simply an artifact of our ability to detect even minute differences between cases and controls? Fig.~\ref{fig:Figure_2}B and~\ref{fig:Figure_2}D show that this was not the case. The median effect size declined only moderately with the number of associations, and most associations corresponded to about a factor of two difference in the taxon abundance. Thus, spurious associations are not weak and could not be discarded based on their effect size.  

\subsection*{Direct association analysis~(DAA)}
\begin{figure}[H]
\begin{center}
\includegraphics[width = 5.5in]{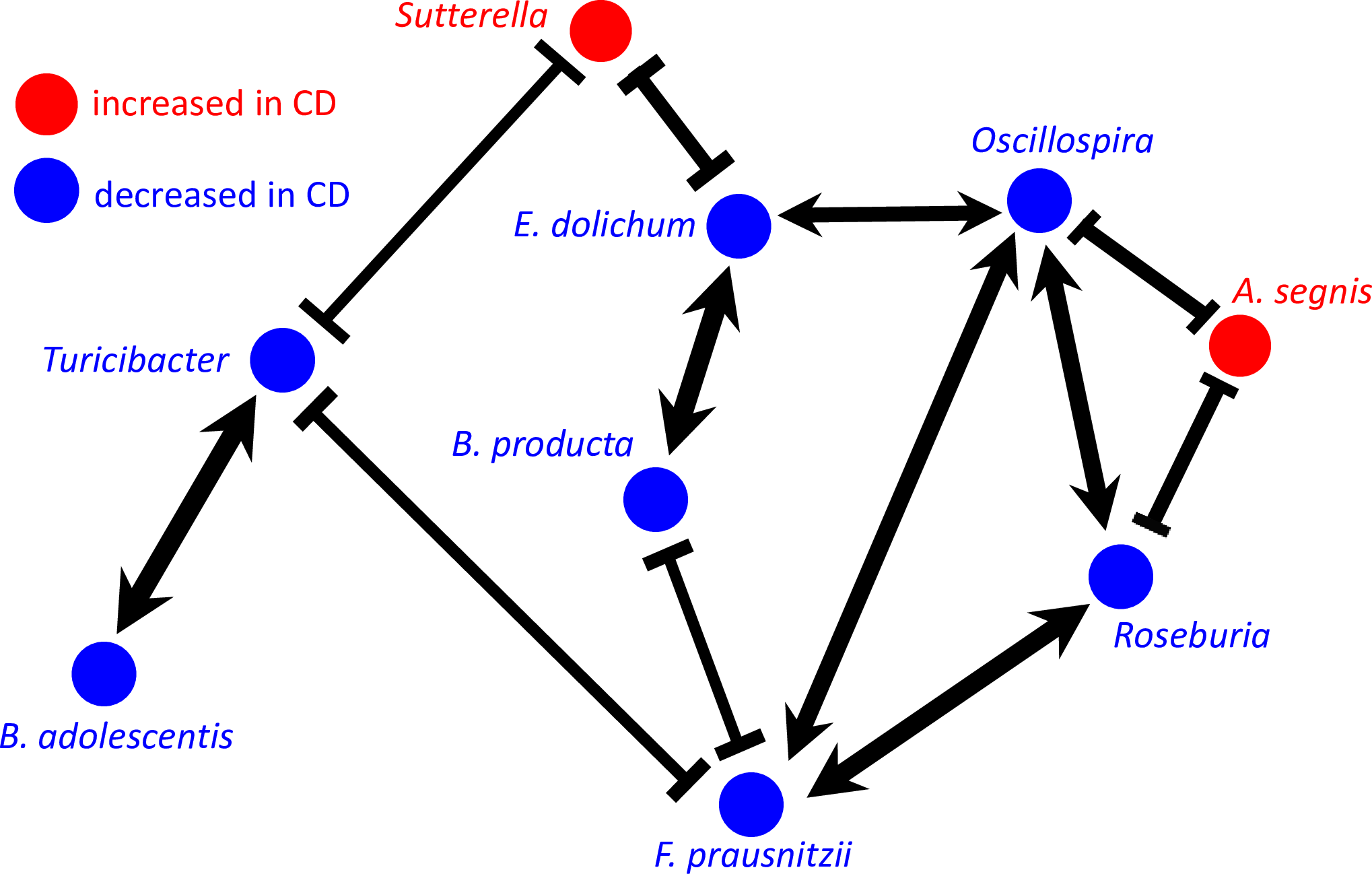}
\vspace{.25in}
\caption{{\bf Network of direct associations with Crohn's Disease.} 
Five species and four genera were found to be significantly associated with Crohn's Disease~($q<0.05$) after correcting for microbial interactions~(Figs. \ref{fig:lognormal} and~\ref{fig:direct_associations}). The links correspond to significant interactions~($q<0.05$) between the taxa with~$J_{ij}>0.27$ or~$J_{ij}<-0.15$; the width of the arrows reflects the strength of the interactions. For comparison, the correlation-based network for directly associated taxa is shown in Figs.~\ref{fig:correlation_matrix} and~\ref{fig:interaction_matrix}, and a complete summary of correlations and interactions for all species pairs is provided in the Supplementary Information.}
	\label{fig:Figure_3}
\end{center}
\end{figure}

Fortunately, the maximum entropy model provides a straightforward way to separate direct from indirect associations. Since direct effects are encoded in~$h$, MWAS should be performed on~$h$ rather than on~$l$. This simple change in the statistical analysis correctly recovered~4 out of~6 directly associated taxa in the synthetic data and yielded no indirect associations even for large cohorts~(Fig.~\ref{fig:Figure_2}A and ~Fig. \ref{fig:Figure2_full}). Similarly good performance was found for the two other synthetic data sets~(~Fig. \ref{fig:alternative_synthetic_effects1}). For the IBD data, DAA also identified a much smaller number of associations compared to traditional MWAS analysis and showed clear saturation at large sample sizes~(Fig.~\ref{fig:Figure_2}B). Direct associations with IBD are summarized in Fig.~\ref{fig:Figure_3} at the genus and species levels, and the entire phylogenetic tree of direct associations is shown in ~Fig. \ref{fig:direct_associations} and in the Supplementary Information.

\begin{figure}[H]
\begin{center}
\includegraphics[width = 6.5in]{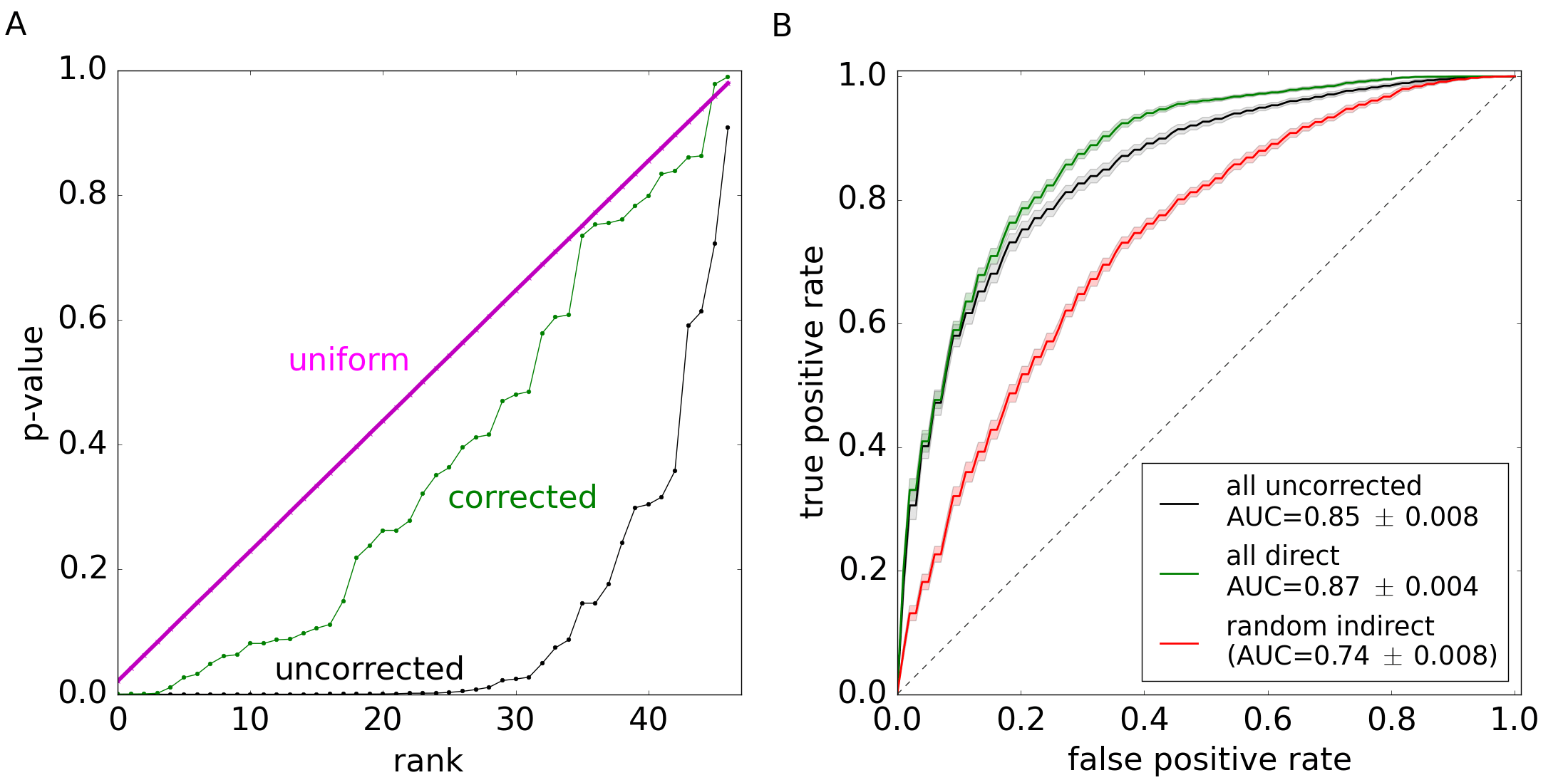}
\vspace{.25in}
\caption{{\bf Direct associations analysis corrects p-value inflation and retains diagnostic accuracy.} \textbf{(A)} The distribution of p-values in DAA closely follows the expected uniform distribution. Because conventional MWAS does not correct for microbial interactions, it yields an excess of low p-values, which is a strong signature of indirect associations. For both methods, p-values were computed using a permutation test. The expected uniform distribution was obtained by sampling from a generator of uniform random numbers. The ranked plot of p-values visualizes their cumulative distribution functions; this is a variant of a Q-Q~plot. \textbf{(B)} Direct associations are a small subset of all associations with IBD (see~Fig. \ref{fig:direct_associations}), yet they retain full power in classifying samples as cases or controls. In contrast, the classification power is substantially reduced for an equally-sized subset of randomly-chosen indirect associations. In each case, we used sparse logistic regression to train a classifier on 80\% of the data and tested its performance on the remaining~20\%~(Methods). The shaded regions show one standard deviation obtained by repeated partitioning the data into training and validation sets. Identical results were obtained with a random forest~\cite{ho:random_forest, breiman:random_forest} and support vector machine~\cite{cortes:svm} classifiers~(Fig.~\ref{fig:other_classifiers})}
\label{fig:Figure_4}
\end{center}
\end{figure}

In addition to associations, DAA also infers the network of direct microbial interactions~(Fig.~\ref{fig:Figure_3},~Figs. \ref{fig:interaction_matrix} and~\ref{fig:interactions_comparison}). While the sample size is insufficient to accurately infer the interactions between every pair of microbes, strong interactions and the overall properties of the interaction network can nevertheless be determined from the data. The interactions inferred by DAA describe only direct effects of the species on each other and do not include induced correlations present in the correlation matrix. That is, DAA controls for the fact that species A and C could be correlated because both interact with species B, but not with each other~(Fig.~\ref{fig:Figure_1}A). The ability of maximum entropy models to separate direct from indirect interactions has been the primary reason for their applications to biological data~\cite{bialek:maxent_neurons, volkov:interactions, mora:maxent_flocks, morcos:direct, chakraborty:hiv_sectors}. Similar to these previous studies, many direct interactions reported in Fig.~\ref{fig:Figure_3} are also present in the correlation-based network, but DAA removes some induced interactions and identifies a few interactions that are not evident in the correlation data; see Figs.~\ref{fig:interaction_matrix} and~\ref{fig:interactions_comparison}. Overall, the interaction network is much sparser than the correlation network in Fig.~\ref{fig:Figure_1}B. In the Supplementary Information, we also compare the results from DAA and SparCC~\cite{friedman:sparcc}, a widely used package to infer correlation networks from microbiome data~(Fig.~\ref{fig:interactions_comparison}).
\vspace{.25in}
\vspace{-.25in}
To demonstrate that DAA isolates direct effects from collective changes in the microbiota, we examined the p-value distribution in this method. The distribution of p-values is commonly used as a diagnostic tool to test whether a statistical method is appropriate for the data. In the absence of any associations, p-values must follow a uniform distribution because the null hypothesis is true~\cite{benjamini_hochberg:fdr}. A few strong deviations from the uniform distribution signal true associations~\cite{storey:q_value}. In contrast, large departures from the uniform distribution typically indicate that the statistical method does not account for some properties of the data, for example, population stratification in the context of genome wide association studies~\cite{mwas:gwas, voorman:p_value_inflation}. Figure~\ref{fig:Figure_4}A compares the distribution of p-values for DAA and a conventional method in MWAS. Consistent with our hypothesis that interspecific interactions cannot be neglected, conventional analysis generates an excess of low p-values and, as a result, a large number of potentially indirect associations. In contrast, the distribution of p-values from DAA matches the expected uniform distribution and, thus, provides strong support for our method.

Finally, we show that indirect associations excluded by DAA do not affect the predictive power of microbiome data. Supervised machine learning such as random forest~\cite{ho:random_forest, breiman:random_forest}, support vector machine~\cite{cortes:svm}, and sparse logistic regression~\cite{walker:logistic_regression, cox:logistic_regression,tibshirani:lasso} were used to classify samples as cases or controls based on their microbiota profile. We found good and identical performance of the classifiers trained either on all taxa detected by conventional MWAS or on a much smaller subset of direct associations detected by DAA~(Fig.~\ref{fig:Figure_4}B). Moreover, the DAA-based classifier showed significantly better performance compared to a classifier trained on an equal number of randomly-selected indirect associations~(Fig.~\ref{fig:Figure_4}B). Thus, DAA reduces the number of associations without losing any information on the disease status and selects taxa with the greatest potential to distinguish health from disease; see Methods for a comparison with the features selected by sparse logistic regression.

\section*{Discussion}
The primary goal of MWAS is to guide the study of disease etiology by detecting microbes that have a direct effect on the host. These direct effects could be very diverse and include secretion of toxins, production of nutrients, stimulation of the immune system, and changes in mucus and bile~\cite{ibd:gwas, xavier:ibd_review_2007}. In addition to the host-microbe interactions, the composition of microbiota is also influenced by the interspecific interactions among the microbes such as competition for resources, cross-feeding, and production of antibiotics~\cite{coyte:stability, foster:cooperation_gut, flint:interactions, bashan:universality, faust:microbiome_interactions, magnusdottir:agora, chu:gut_antibiotics, riley:warfare, czaran:warfare, dethlefsen:assembly_review, mackie:interactions_gut}. In the context of MWAS, microbial interactions contribute to indirect changes in microbial abundances, which are less informative of the disease mechanism and are less likely to be valuable for follow-up studies or in interventions. Here, we estimated the relative contribution of indirect associations to MWAS and showed how to isolate direct from indirect associations.

Our main result is that interspecific interactions are sufficiently strong to generate detectable changes in the abundance of many microbes that are not directly linked to host phenotype. As a result, conventional approaches to MWAS detect a large number of spurious associations and produce inflated p-values that do not match their expected distribution~(Fig.~\ref{fig:Figure_4}A). These challenges are resolved by Direct Association Analysis~(DAA), which uses maximum entropy models to explicitly account for interspecific interactions. We applied DAA to a large data set of pediatric Crohn's disease and found that it restores the distribution of p-values and substantially simplifies the pattern of dysbiosis while retaining full classification power of a conventional MWAS.  

The relatively simple dysbiosis identified by DAA in IBD has strong support in the literature and offers interesting insights into disease etiology. Four of the taxa identified by our method have a well-established role in IBD: \textit{B. adolescentis}, \textit{F. prausnitzii}, \textit{B. producta}, and~\textit{Roseburia}. They have been repeatedly found to have lower abundance in both Crohn's disease and ulcerative colitis~\cite{morgan:roseburia_down, machiels:butyrate_down, morgan:ibd_metagenomics, travis:roseburia_genome, forbes:gut_inflammatory_diseases, joossens:prausnitzii_low, sokol:prausnitzii_low, blautia:reduced_butyrate}, and several studies have demonstrated their ability to suppress inflammation and alleviate colitis~\cite{sokol:prausnitzii_anti_inflammatory, zhang:faecalibacterium_immune, qiu:faecalibacterium_immune, forbes:inflammatory_diseases, scharek:bifidobacterium_immune, travis:roseburia_genome}. \textit{Bifidobacterium} species occupy a low trophic level in the gut and ferment complex polysaccharides such as fiber~\cite{oyetayo:bifidobacterium_review, duranti:bifidobacterium_genome}. Fermentation products include lactic acid, which promotes barrier function, and maintains a healthy, slightly acidic environment in the colon~\cite{bifidobacterium:lactic}. Due to these properties \textit{Bifidobacterium} species are commonly used as probiotics~\cite{oyetayo:bifidobacterium_review}. \textit{F. prausnitzii}, \textit{Blautia producta} and~\textit{Roseburia} occupy a higher trophic level and ferment the byproducts of polysaccharides digestion into short-chain fatty acids (SCFA), which are an important energy source for the host~\cite{louis:scfa_review, travis:roseburia_genome, morgan:ibd_metagenomics, nigel:butyrate}. 

The ability of DAA to detect taxa strongly associated with IBD is reassuring, but not surprising. What is surprising is that many strong associations are classified as indirect by our method. For example,~\textit{Roseburia} and~\textit{Blautia} are the only genera of~\textit{Lachnospiraceae} that DAA finds to be directly linked to the disease. In sharp contrast, traditional MWAS report seven genera in this family that are strongly associated with IBD~\cite{wang:risk}. All seven genera are involved in SCFA metabolism, but their specializations differ. Species in~\textit{Blautia} genus are major producers of acetate, a SCFA that is commonly involved in microbial crossfeeding~\cite{carbonero:hydrogen, louis:cancer_microbes}. In particular, many species extract energy from acetate by converting it into butyrate, another SCFA that plays a major role in gut health by nourishing colonocytes and regulating the immune function~\cite{louis:cancer_microbes,louis:scfa_review}. \textit{Roseburia} genus specializes almost exclusively in the production of butyrate and acts as a major source of butyrate for the host~\cite{louis:scfa_review, kettle:roseburia}. Thus, our findings suggest that butyrate production plays an important role in IBD etiology and that the dysregulation of this process is directly linked to the depletion of~\textit{Roseburia} and possibly~\textit{Blautia}. 

The important role of butyrate is further supported by our detection of~\textit{E. dolichum} and~\textit{Oscillospira}, which are known to produce butyrate~\cite{eeckhaut:butyrate_edolichum, louis:butyrate_edolichum, oscillospira_butyrate}. The latter taxon has not been detected in three independent analyses of this IBD data set~ \cite{gevers:risk, risk:followup, wang:risk} presumably because its involvement is masked by indirect associations and interactions with other microbes. Several other studies support this DAA finding and confirm that~\textit{Oscillospira} is suppressed in IBD \cite{oscillospira_reduced, oscillospira_reduced1}. \textit{Oscillospira} was also found to be positively associated with leanness and negatively associated with the inflammatory liver disease \cite{oscillospira_lean, oscillospira_leantwin, oscillospira_liver}. The interactions between~\textit{Oscillospira} and the host appears to be quite complex and involve the consumption of host-derived glycoproteins including mucin, production of SCFA, and modulation of bile-acid metabolism \cite{bileacid_Cdiff, oscillospira_butyrate}. The latter interaction was suggested to be a major factor in the protective role of~\textit{Oscillospira} against infections with~\textit{Clostridium difficile} \cite{bileacid_Cdiff, oscillospira_Cdiff1,oscillospira_Cdiff}.

The final taxon that was suppressed in IBD is \textit{Turicibacter}. This genus is not very well characterized, and few MWAS studies point to its involvement in IBD~\cite{gevers:risk, wang:risk, minamoto:turicibacter_dogs}. Two studies in animal models, however, directly looked into the connection between IBD and~\textit{Turicibacter}~\cite{werner:iron_turicibacter, presley:turicibacter_immunity}. The first study found that iron limitation eliminates colitis in mice while at the same time restoring the abundance of \textit{Turicibacter}, \textit{Bifidobacterium}, and four other genera~\cite{werner:iron_turicibacter}. The second study identified \textit{Turicibacter} as the only genus that is fully correlated with immunological differences between mice resistant and susceptible to colitis: high abundance of~\textit{Turicibacter} in the colon predicted high levels of MZ B and iNK T cells, which are potent regulators of the immune response~\cite{presley:turicibacter_immunity}. Moreover, \textit{Turicibacter} was the only genus positively affected by the reduction in CD$8^{+}$~T cells. Thus, our method identified a taxon that is potentially directly linked to IBD via the modulation of the immune system.

Perhaps the most unexpected finding was our detection of~\textit{A. segnis} and~\textit{Sutterella} as the only species and genus increased in disease compared to 26 positive associations detected by the previous analysis~\cite{wang:risk}. All other associations were classified as indirect even though they often corresponded to much more significant changes in abundance between IBD and control groups. Thus, our results indicate that expansion of many taxa including opportunistic pathogens is driven by their interactions with the core IBD network shown in Fig.~\ref{fig:Figure_3}. One possibility is that the dysbiosis of the symbiotic microbiota makes it less competitive against other bacteria and opens up niches that can be colonized by opportunistic pathogens. The other, less explored possibility, is that commensal microbiota can not only protect from pathogens, but also facilitate their invasion, a phenomenon that has been recently demonstrated in bees~\cite{schwarz:bee_pathogen}.

Little is known about the specific roles that~\textit{A. segnis} and~\textit{Sutterella} play in IBD, and more generally in gut health.~\textit{Aggregatibacter} is a common member of the oral microbiota that thrives in local infections such as periodontal disease and bacterial vaginosis \cite{aggregatibacter_periodontal, segnis_periodontal, aggregatibacter_vaginosis}. The high abundance of \textit{A. segnis} is also associated with an increased risk of IBD recurrence \cite{segnis_IBD}. \textit{Sutterella}, on the other hand lacks overt pathogenicity, and MWAS produced inconsistent findings~\cite{lavelle_sutterella_ibd, mangin:molecular_inventory_cd, gophna:differences, tyler:sutterella_pouch, hansen:biscuit, hiippala:sutterella_phenotype, mukhopadhya:sutterella_ibd} on its involvement in IBD. Some studies reported that~\textit{Sutterella} is increased in patients with good outcomes~\cite{gevers:risk, tyler:sutterella_pouch} while other studies found positive or no association between~\textit{Sutterella} and IBD~\cite{wang:risk, hiippala:sutterella_phenotype, hansen:biscuit, mukhopadhya:sutterella_ibd, mangin:molecular_inventory_cd}. Experimental investigations showed that \textit{Sutterella} lacks many pathogenic properties; in particular, it does not induce a strong immune-response and has only moderate ability to adhere to mucus~\cite{hiippala:sutterella_phenotype, mukhopadhya:sutterella_ibd}. Further, \textit{Sutterella} strains from IBD and control patients showed no phenotypic differences in metabolomic, proteomic, and immune response assays~\cite{mukhopadhya:sutterella_ibd}. Nevertheless, \textit{Sutterella} is strongly associated with worse behavioral scores in children with autism spectrum disorder and Down syndrome~~\cite{williams:sutterella_autism, wang:sutterella_autism, biagi:sutterella_down_syndrome}. Therefore, the direct link between~\textit{Sutterella} and IBD could involve the gut-brain axis.
 
In summary, we found a small number of taxa can explain extensive dysbiosis in IBD and accurately predict disease status. Directly associated taxa have strains with dramatically different abilities to trigger colitis and are specifically targeted by the immune system of patients and animals with IBD~\cite{palm:antibody_coating}. Previous studies of these taxa point to facilitated colonization by pathogens, butyrate production, immunomodulation, bile metabolism, and the gut-brain axis as the primary factors in the etiology of IBD.

Many disorders are accompanied by substantial changes in host microbiota, but our work shows that only a small subset of these changes could be directly related to the disease. Similarly, only a handful of taxa could drive the dynamics of the ecosystem-level changes in the environment. To untangle the complexity of such dysbioses, it is important to account for microbial interactions using mechanistic or statistical methods. Direct association analysis proposed in this paper is a simple statistical approach based on the principle of maximum entropy. DAA can be applied to any microbiome data set that is sufficiently large to infer interspecific interactions.

\section*{Methods}
The data used in this study was obtained from Ref.~\cite{gevers:risk}, which reported changes in the microbiome of newly-diagnosed, treatment-naive children with IBD compared to controls. This data was recently analyzed in Ref.~\cite{wang:risk}, and we followed all the statistical procedures adopted in that study to enable direct comparison of the results. Specifically, we used a permutation test on mean log-transformed abundances to determine the statistical significance of an association. 

To fit the maximum entropy model to the data, we first computed the mean log-abundance for each genus~$m_i$ and the covariance in the log-transformed abundances~$C_{ij}$. The interaction matrix was computed as~$J=C^{-1}$ by performing singular value decomposition~\cite{stewart:early_svd} and removing all singular values that were comparable to the amount of noise present in the data. The host effects were computed as~$h=Jm$. See Supplementary Information for further details.

All computation was carried out in Python environment. We used \texttt{scikit-learn~0.15.2}~\cite{pedregosa:scikit} for hierarchical clustering and to build the supervised classifiers used in Fig.~\ref{fig:Figure_4}B of the main text and ~Fig. \ref{fig:other_classifiers}. The variance in the accuracy of classification was evaluated through 5-fold stratified cross-validation with 100~random partitions of the data into the training and validation sets.  For all findings, statistical significance was evaluated with Fisher's exact test~(permutation test) with~$10^6$ permutations. False discovery rate was controlled to be below~5\% following Benjamini-Hochberg procedure~\cite{benjamini_hochberg:fdr}.

For sparse logistic regression, we confirmed that the penalty parameter was in the range where the results are insensitive to its specific value. The features selected by this classifier in Fig.~\ref{fig:Figure_4} are as follows: \textit{Erysipelotrichales}, \textit{Pasteurellales}, \textit{Turicibacterales}~(also significant in DAA), and \textit{Enterobacteriales}~(not significant in DAA) at the order level; \textit{Clostridiaceae} and \textit{Pasteurellaceae}~(also significant in DAA) and \textit{Enterobacteriaceae} and \textit{Erysipelotrichaceae}~(not significant in DAA) at the family level; \textit{Roseburia}~(also significant in DAA) and \textit{Dialister}, \textit{Aggregatibacter}, and \textit{Haemophilus}~(not significant in DAA) at the genus level; and \textit{B.~adolescentis}, \textit{F.~prausnitzii}, and \textit{E.~dolichum}~(also significant in DAA) and \textit{Prevotella~copri} and \textit{Haemophilus~parainfluenzae}~(not significant in DAA) at the species level. In total, both DAA and the sparse logistic regression relied on~17 features with~9 of them being the same. Thus, DAA identified many features that were also selected by the machine learning algorithm for their predictive value. At the same time, the results of DAA and the sparse logistic regression were not exactly the same and, therefore, could be complementary to each other.

\nolinenumbers
\newpage
\renewcommand{\thefigure}{S\arabic{figure}}
\renewcommand{\thesection}{S\arabic{section}}
\renewcommand{\thetable}{S\arabic{table}}

\renewcommand\Re{\operatorname{Re}}
\renewcommand\Im{\operatorname{Im}}
\setcounter{figure}{0}

\section*{Supplementary information}
\textbf{\large{Model of community composition}}\\ Here we describe a mathematical model of community composition, that we use to correct for microbial interactions in microbiome-wide association studies.

\begin{figure}
\begin{center}
\includegraphics[width = 6in]{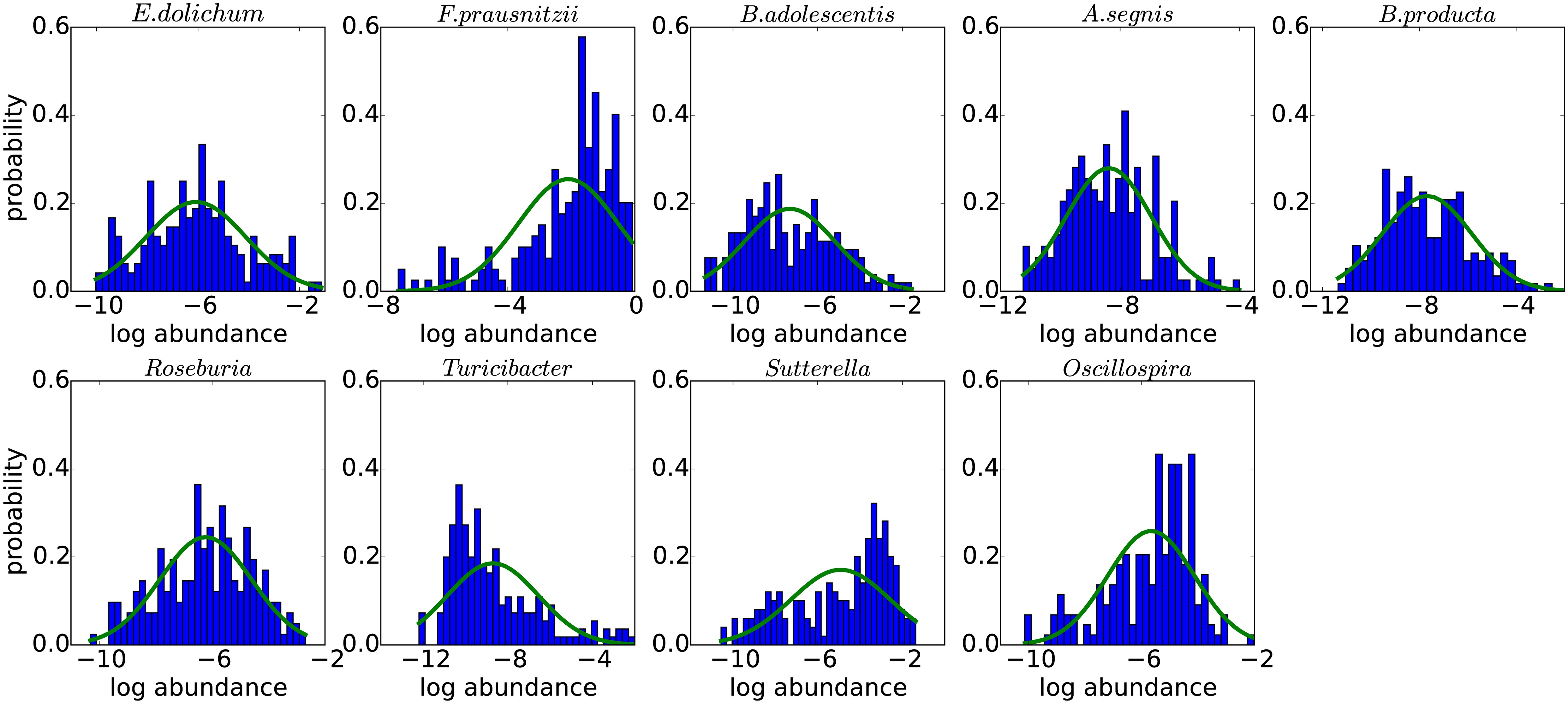}
\vspace{.25in}
\caption{{\bf Microbial abundances follow the log-normal distribution.} The histograms show probability distributions of the relative log-abundance for the species and genera detected by DAA (summarized in Fig.~\ref{fig:Figure_3}). The best fit of a Gaussian distribution is shown in green.}
\label{fig:lognormal}
\end{center}
\end{figure}

\textit{Log-transformation of abundances}\\
The environment within a host is constantly changing due to variations in diet, immune response, phage activity and other factors. As a result, microbial growth rates should be highly variable and produce multiplicative fluctuations in the community composition, which are better captured on logarithmic rather than on linear scale. Indeed, the abundances of many gut species follow a log-normal distribution (Fig.~\ref{fig:lognormal}), and recent work shows that a log-transformation of abundances increases the power and quality of microbiome studies~\cite{wang:risk}. Therefore, we chose to carry out all of the analysis and modeling on natural logarithms of relative abundances computed with a pseudocount of one read. For simplicity, we refer to these quantities as abundances in the following and denote them as~$l_{i}$ with the subscript identifying the species under consideration.

\textit{Maximum entropy models}\\
Microbiota composition is highly variable among people in both health and disease~\cite{wang:risk} and needs to be described via a multivariate probability distribution~$P(\{l_i\})$. The amount of data in a large microbiome-wide association study, however, is sufficient to reliably determine only the first and second moments of~$P(\{l_i\})$. This situation is common in the analysis of biological data and has been successfully managed with the use of maximum entropy distributions~\cite{sander:maxent_review}. These distributions are chosen to be as random as possible under the constraints imposed by the first and second moments. Maximum entropy models introduce the least amount of bias and reflect the tendency of natural systems to maximize their entropy. In other contexts, these models have successfully described the dynamics of neurons~\cite{bialek:maxent_neurons}, forests~\cite{volkov:interactions}, and flocks~\cite{mora:maxent_flocks}, and even predicted protein structure~\cite{morcos:direct} and function~\cite{chakraborty:hiv_sectors}. In the context of microbiomes, a recent work derived a maximum entropy distribution for microbial abundances using the principle of maximum diversity~\cite{fisher:habitat_fluctuations}.  

Let us denote abundance means and covariances computed from the data by the vector~$m$ and matrix~$C$ respectively. The constraints on the maximum entropy distribution are then expressed as

\begin{equation}
\begin{aligned}
& \langle l_{i} \rangle = m_i \\
& \langle l_{i}l_{j} \rangle - \langle l_{i} \rangle \langle l_{j} \rangle = C_{ij}
\end{aligned}
\label{eq:constraints}
\end{equation}

and the maximum entropy distribution takes the following form

\begin{equation}
P(\{l_{i}\}) = \frac{1}{Z} e^{\sum_{i}h_{i}l_{i} + \frac{1}{2}\sum_{ij}J_{ij}l_{i}l_{j}}
\label{eq:maxent_SI}
\end{equation}

\noindent which is similar to the Ising model of statistical physics, but with continuous rather than discrete degrees of freedom. The variables~$h_{i}$ and~$J_{ij}$ arise as Lagrange multipliers for the first and second moment constraints during entropy maximization. In statistical physics, they describe local magnetic fields that align spins~$l_{i}$ and interactions between spins~$l_{i}$ and~$l_j$. The constant~$Z$, known as the partition function, ensures that the distribution is normalized:

\begin{equation}
Z = \int\prod_{i}dl_{i}e^{\sum_{i}h_{i}l_{i} + \frac{1}{2}\sum_{ij}J_{ij}l_{i}l_{j}}
\label{eq:Z_definition}
\end{equation}

\noindent Note that~$Z$ is a multi-dimensional Gaussian integral.

\textit{Host effects vs. species interactions}\\
To interpret this maximum entropy distribution in terms of biologically relevant factors such as microbial interactions and properties of the host, we can rewrite equation~(\ref{eq:maxent_SI}) as follows

\begin{equation}
P(\{l_{i}\}) = \frac{1}{Z} e^{\sum_{i}H_{i}l_{i}}
\label{eq:maxent_rewritten}
\end{equation}

\noindent where

\begin{equation}
H_{i} = h_{i} + \frac{1}{2}\sum_{j}J_{ij}l_{j}
\label{eq:H_definition}
\end{equation}

\noindent describe the quality of the local environment for species~$i$: the higher~$H_{i}$, the more abundant the species. The quality of the environment can be decomposed into external variables such as temperature or metabolite concentrations~$V_{\alpha}$ and the species' response to these variables~$R_{i\alpha}$ as

\begin{equation}
H_{i} = \sum_{\alpha} R_{i\alpha} V_{\alpha}
\label{eq:V_definition}
\end{equation}

\noindent We can further decompose the external variables~$V_{\alpha}$ into host factors~$V_{\alpha}^{h}$ and influences of other species, e.g., due to metabolite secretion or production of antibiotics: 

\begin{equation}
V_{\alpha} = V_{\alpha}^{h} + \sum_{j} P_{\alpha j} l_{j}
\label{eq:V_decomposition}
\end{equation}

\noindent where~$P_{\alpha j}$ describes the influence of microbe~{$j$} on variable~$\alpha$.

Upon combining equations~(\ref{eq:V_definition}) and~(\ref{eq:V_decomposition}), we can express~$H_{i}$ as

\begin{equation}
H_{i} = \sum_{\alpha}R_{i\alpha}V_{\alpha}^{h} + \sum_{\alpha j}R_{i\alpha}P_{\alpha j}l_{j}
\label{eq:H_decomposition}
\end{equation}

Comparison of this equation to equation~(\ref{eq:H_definition}) shows that we can identify~$h_{i}=\sum_{\alpha}R_{i\alpha}V_{\alpha}$ with the direct effects of the host and~$J_{ij}=2\sum_{\alpha}R_{i\alpha}P_{\alpha j}$ with the interactions among the microbes. 

\textbf{\large{Inference of model parameters}}

Here we describe the procedure of learning the parameters of the maximum entropy model from the data. Our approach closely follows that of Refs.~\cite{sander:maxent_review},~\cite{morcos:direct} and~\cite{chakraborty:hiv_sectors}.

\textit{Relating $h$ and $J$ to $m$ and $C$}\\
To infer model parameters~$h_{i}$ and~$J_{ij}$, we need to relate them to empirical observations such as the means and covariances of the abundances. These relationships can be conveniently obtained from the derivatives of the partition function, which is the standard approach in statistical physics. Indeed, the mean abundances can be expressed as

\begin{equation}
\langle l_{k} \rangle = \frac{1}{Z} \int\prod_{i}dl_{i}e^{\sum_{i}h_{i}l_{i} + \frac{1}{2}\sum_{ij}J_{ij}l_{i}l_{j}}l_{k} = \frac{\partial \ln Z}{\partial h_k}
\label{eq:Z_l}
\end{equation}

\noindent A similar relationship holds for the covariance matrix:

\begin{equation}
\langle l_{i}l_{j} \rangle - \langle l_{i} \rangle \langle l_{j} \rangle =  \frac{\partial^2 \ln Z}{\partial h_i \partial h_j}
\label{eq:Z_C}
\end{equation}

To complete the calculation, we need to compute the partition function defined by equation~(\ref{eq:Z_definition}). The result reads

\begin{equation}
Z = \frac{1}{\sqrt{\det(J/2\pi)}}e^{\frac{1}{2}h^{T}J^{-1}h}
\label{eq:Z_computed}
\end{equation}

where symbols without indexes are treated as vectors or matrices.

From equation~(\ref{eq:Z_computed}), we immediately find that

\begin{equation}
\begin{aligned}
& m = J^{-1} h\\
& C = J^{-1}
\end{aligned}
\label{eq:mC_computed}
\end{equation}

\noindent which can be inverted to obtain

\begin{equation}
\begin{aligned}
& h = C^{-1} m\\
& J = C^{-1}
\end{aligned}
\label{eq:hJ_computed}
\end{equation}

\textit{Inverting the covariance matrix}\\
It is clear from equation~(\ref{eq:hJ_computed}) that the key step in obtaining the model parameters is the inversion of the covariance matrix. However, this matrix is likely to be degenerate or ill-conditioned because of the insufficient amount of data or very strong correlations between microbial abundances. To overcome this difficulty, we computed a pseudoinverse of~$C$ as described in the following sections. Briefly, we used singular value decomposition~\cite{stewart:early_svd} of~$C$ in terms of two orthogonal matrices~$U$ and~$V$ (since~$C$ is symmetric,~$U=V$) and a diagonal matrix~$\Lambda$:

\begin{equation}
C=U\Lambda V^T
\label{eq:svd}
\end{equation}

\noindent Some diagonal elements of~$\Lambda$ were small and comparable to the levels of noise~(or uncertainty), so we set the corresponding elements of~$\Lambda^{-1}$ to zero. Specifically,~$\Lambda^{-1}_{kk}$ was set to zero for all~$k$ such that~$\Lambda_{kk}<\lambda_{\mathrm{min}}$, where~$\lambda_{min}$ was a predetermined threshold. A regular inverse~($\Lambda^{-1}_{kk}=1/\Lambda_{kk}$) was used for the rest of the elements. The choice of the threshold and the robustness of the results to the variation in~$\lambda_{\mathrm{min}}$ are discussed in the section on data analysis. This procedure ensured that we do not infer large changes in host fields~$h$ due to fluctuations in the estimate of~$\langle l \rangle$. The inverse of~$C$ was then computed as~$C^{-1} = V\Lambda^{-1} U^{T}$, where we used the fact that the inverse of an orthogonal matrix is its transpose. 

\textbf{\large{Origin of spurious associations and Direct Associations Analysis}}

\textit{Microbial interactions introduce spurious associations}\\
In microbiome-wide association studies, we are typically interested in the changes in microbial abundances~$\Delta m$ between two groups of subjects. From equation~(\ref{eq:mC_computed}), we can relate~$\Delta m$ to the changes in the phenotype of the host~$\Delta h$:

\begin{equation}
\Delta m = C \Delta h
\label{eq:spurious_origin}
\end{equation}

\noindent This formula clearly illustrates the origin of spurious associations. Imagine that there is a small number of species directly linked to host phenotype, i.e.~$\Delta h$ is a sparse vector. Because~$C$ is a dense matrix~(see Fig.~1b in the main text), equation~(\ref{eq:spurious_origin}) predicts that~$\Delta m$ is dense, i.e. the abundances of most species are affected. The sizes of these effects are variable and depend on the magnitude of the off-diagonal elements of~$C$. Except for the strongly interacting species, the largest changes in~$m$ are likely to mirror the largest changes in~$h$ and result in significant associations. In large samples, however, smaller effects become detectable that could either reflect small direct effects or the secondary, indirect effects due to microbial interactions. As a result, the number of associations grows with the sample size, and the relationship between associated species and host phenotype becomes obscured. Fig.~2 in the main text presents evidence for a large number of spurious associations in both synthetic and real data. 

\textit{Removing indirect associations}\\
Equation~(\ref{eq:spurious_origin}) offers a straightforward way to correct for microbial interactions and separate direct from indirect associations. Indeed, for each species, we can compute the corresponding change in the host field as

\begin{equation}
\Delta h_{i} = \sum_{j} \left(C^{-1}\right)_{ij} \Delta m_j
\label{eq:daa}
\end{equation}

\noindent The statistical significance of this change can be determined via the permutation test followed by the Benjamini-Hochberg procedure to correct for multiple hypothesis testing~\cite{benjamini_hochberg:fdr}.

\textbf{\large{Assumptions and limitations of DAA}}

\textit{Pairwise interactions are sufficient}\\
So far, we have considered only pairwise interactions between the taxa. This is a common assumption in maximum entropy models, which reflects the need for very large data sets to reliably infer higher-order interactions~\cite{sander:maxent_review,bialek:maxent_neurons, volkov:interactions, mora:maxent_flocks, morcos:direct, chakraborty:hiv_sectors}. While fitting higher-order interactions is impractical, we can nevertheless test whether they make a significant contribution to the patterns of co-occurrence observed in IBD data. To this purpose, we computed third and fourth order moments of microbial abundances in IBD data and compared them to the corresponding moments predicted by our maximum entropy model. This is a meaningful test because only the first and second moments were used to fit the model to the data.
		
The predictions of our model follow from the properties of the multivariate Gaussian distribution and can be summarized as follows:

\begin{equation}
\begin{aligned}
& \langle l_{i} l_{j} l_{k} \rangle = m_{i} m_{j} m_{k} + m_{i} C_{jk} +  m_{j}C_{ik} + m_{k} C_{ij} \\
& \langle(l_{i} - \langle l_{i} \rangle)(l_{j} - \langle l_{j} \rangle)(l_{k} - \langle l_{k} \rangle)\rangle = 0\\
& \langle(l_{i} - \langle l_{i} \rangle)(l_{j} - \langle l_{j} \rangle)(l_{k} - \langle l_{k} \rangle)(l_{m} - \langle l_{m} \rangle)\rangle = C_{ij}C_{km}+C_{im}C_{jk}+C_{ik}C_{jm}
\end{aligned}
\label{eq:higher_moment}
\end{equation}

\begin{figure}
\begin{center}
\includegraphics[width = 6.5in]{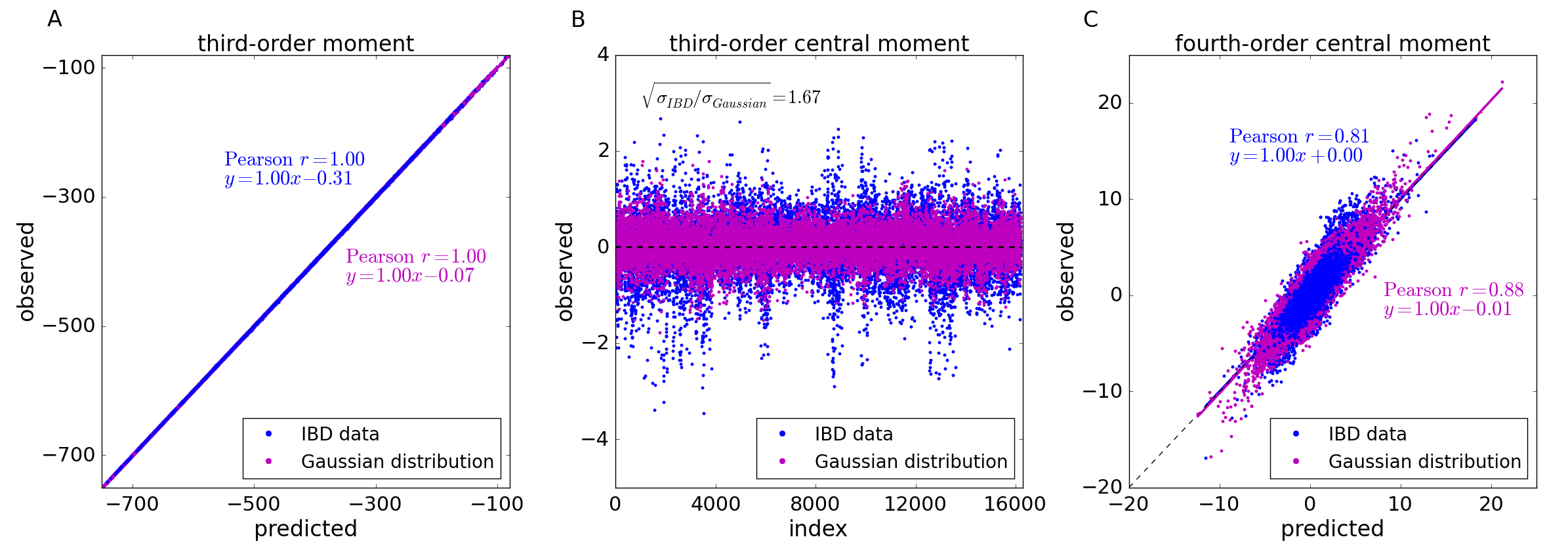}
\vspace{.15in}
\caption{\textbf{Pairwise interactions are sufficient to explain the patterns of microbial co-occurence.} The parameters in our maximum entropy model were chosen to fit only the first and the second moments of the multivariate distribution of microbial abundances. Nevertheless, the model captures most of the higher-order correlations in the data suggesting pairwise interactions are sufficient to accurately describe the patterns of microbial co-occurences. \textbf{(A)} For each choice of three genera, the third order moment was computed by averaging the product of the log-abundances over all the samples in the IBD data (``observed'') or from Eq.~(17)~(``predicted''), which states the predictions of the maximum entropy model. The plot shows excellent agreement between the two quantities. \textbf{(B)} For each choice of three genera~(``index''), we plot the third-order central moment computed from the IBD data~(``observed'') and from an equally-sized sample drawn from our maximum entropy model~(``Gaussian distribution''). The latter quantifies the expected deviations between the observations and predictions due to the finite size of the sample. \textbf{(C)} Same as (A), but for the fourth-order central moment. The expected level of noise is quantified via a sample from the maximum entropy model that obeys Eq.~(17) exactly in the limit of infinite sample size. The correlation coefficient between ``observed'' and ``predicted'' values from this sample sets the upper bound on the expected correlation coefficient in IBD data.}
\label{fig:higher order}
\end{center}  
\end{figure}

\noindent The model predicts that the third central moments vanish, and indeed the corresponding values in the data are close to zero~(Fig.~\ref{fig:higher order}). The observed deviation is consistent with the level of  noise seen in a random Gaussian sample drawn from the maximum entropy distribution;  the size of the sample equaled that of the IBD data. Further, the predictions for the non-central moments are highly correlated with the moments observed in IBD data~(Fig.~\ref{fig:higher order}) with Pearson's~$r$ equal to 1 and 0.81 for third and fourth moments respectively. The deviations of~$r$ from~$1$ are largely due to the uncertainty in the values of the observed moments. Indeed, we obtained~$r=1$ and~$r=0.88$ for the correlation between predicted and observed third and fourth order moments for the random sample drawn from our maximum entropy distribution. Since the higher moments of the maximum entropy distribution satisfy Eq.~(\ref{eq:higher_moment}) exactly, the observed values of~$r$ set the upper bound on the correlation coefficient that can be obtained given the sample size in the IBD data set.

\begin{figure}
\begin{center}
\includegraphics[width = 3.7in]{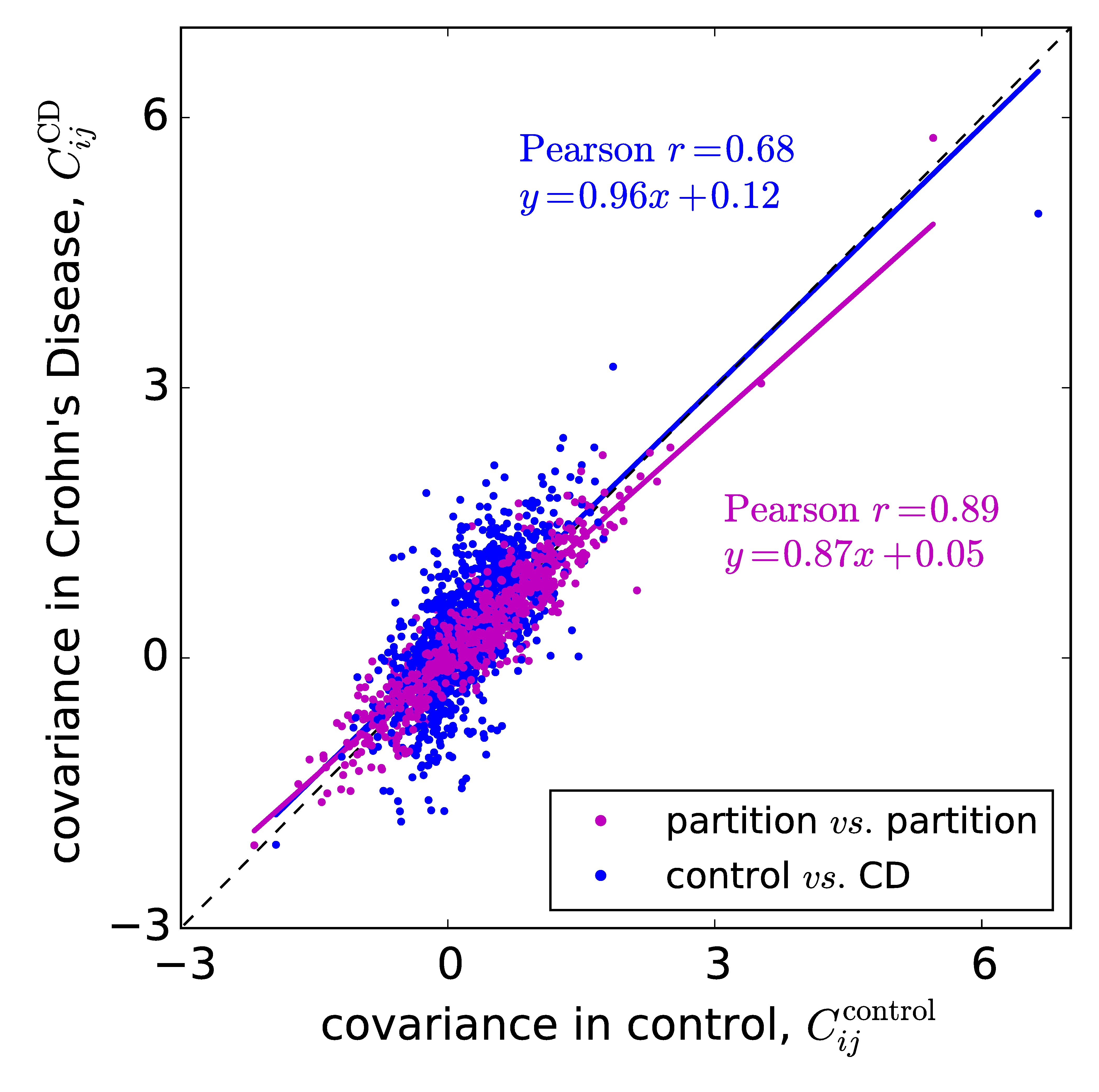}
\caption{\textbf{Microbial interactions are only weakly affected by host phenotype.} To determine whether Crohn's disease drastically alters the pattern of microbial interactions, we computed and compared the covariance matrixes~$C^{\mathrm{CD}}$ and~$C^{\mathrm{control}}$ for CD and control groups respectively. The results of this calculation for IBD data are shown in blue. Each dot corresponds to a matrix element of $C_{ij}$, which is the covariance between the log-abundances of genera i and j. The $x$-coordinate is the covariance computed in the control group and the $y$-coordinate is the covariance computed in the CD group. To estimate the expected level of noise, we carried out the same analysis on two random partitions of the data that contain both controls and subjects with CD~(shown in magenta). Since the groups are drawn from the same distribution, their covariance matrices must be identical on average. The spread of the magenta data points, therefore, sets the upper limit on the correlation coefficient between~$C^{\mathrm{CD}}$ and~$C^{\mathrm{control}}$. We note, however, that this upper bound is unlikely to be reached for IBD data because some taxa have different noise levels in CD and control groups: eg. the taxa depleted in CD have a low abundance in this group and, therefore, higher error in the estimates of the correlation coefficients with other taxa. Overall, both IBD and partitioned data lie close to the diagonal and exhibit similar levels of variation. Thus, using the same covariance matrix for both CD and control groups is a reasonable first approximation. This approximation is valuable because it reduces the uncertainty in~$C_{ij}$ by allowing us to use the entire data to compute covariances and because it improves the stability of DAA to errors in~$C$ (see Fig. \ref{fig:dh_subsampled}).}
\label{fig:cov_assumptions}
\end{center}  
\end{figure}

\textit{Host phenotype affects~$h$, but not~$J$}\\
An important assumption behind Eq.~(\ref{eq:daa}) is that the interspecific interactions are not affected by host phenotype, i.e.~$C$ and~$J$ are the same for control and disease groups. Deviations from this assumption are certainly possible, but they represent higher order effects, which are absent in a simple linear-response model of microbial communities given by Eq.~(\ref{eq:H_decomposition}). Moreover, current sample sizes are insufficient to accurately infer and compare the covariance matrices for each of the groups. Association tests between microbial interactions and host phenotype are further complicated by the large number of interspecific interactions, which leads to a severe reduction in statistical power. Therefore, we did not attempt to identify specific interactions that are affected by IBD; instead, we assessed the overall similarity between the covariance matrices~$C^{\mathrm{CD}}$ and~$C^{\mathrm{control}}$ computed for patients with and without Crohn's disease~(Fig.~\ref{fig:cov_assumptions}). We found that the plot of the matrix elements of $C^{\mathrm{CD}}$ vs. $C^{\mathrm{control}}$ clustered around the diagonal with the coefficient of linear regression equal to 0.96, suggesting that the structure of correlations is similar for the two phenotypic groups. The spectral properties of the matrices are also similar. \\
To perform a more quantitative comparison we also computed the Pearson correlation coefficient between the matrix elements of $C^{\mathrm{CD}}$ vs. $C^{\mathrm{control}}$ ($r=0.7$). However, interpreting the value of the correlation coefficient is non-trivial because it is very sensitive to the noise in the data and the uncertainty in the individual matrix elements is high, especially for taxa with low abundance. One way to estimate the expected level of noise is to compare the observed correlation coefficient to the correlation coefficient for two subsamples of the shuffled data drawn without preserving the diagnosis labels, but of the same size as the CD and control groups. This coefficient must equal 1 in the limit of infinitely large data, so it sets the upper limit on r that can be observed between $C$ computed for CD and control groups, even when there are no differences in the interactions. We note, however, that this upper bound is unlikely to be reached for IBD data because some taxa have different noise levels in CD and control groups. Indeed, the taxa depleted in CD have a low abundance in this group and, therefore, higher error in the estimates of the correlation coefficients with other taxa. We found that the correlation coefficient $r$ between two random subsets was about $0.9$, suggesting that high level of noise is the likely explanation for the spread of the data away from the diagonal in Fig.~\ref{fig:cov_assumptions}.

\textit{Robustness of inference to the uncertainties in the covariance matrix}\\
Since the sample size in the IBD data set is not sufficient to infer every element of the covariance matrix accurately, it is important to determine how the uncertainty in~$C$ affects DAA results. To this end, we repeatedly subsampled the IBD data set to half of its size and examined the variation in the gross properties of~$C$ and changes in~$h$ and~$\Delta h$. Fig.~\ref{fig:eigenvalues_uncertainty} shows that the eigenvalues of~$C$ are extremely robust and are virtually unaffected by the subsampling of the data. Similarly, there is only small variation in the values of~$\Delta h$ between control and CD groups~(Fig.~\ref{fig:dh_subsampled}). For genera detected by DAA, the values of~$\Delta h$ together their error bars due to subsampling are well outside the region where~$\Delta h$ are expected to lie under the null hypothesis of no association between the genus and Crohn's disease.

\textit{Compositional effects}\\
Microbial abundances are usually normalized by the total number of reads in the sample to eliminate the noise introduced during sample preparation, for example, at DNA extraction and amplification steps. Other normalization schemes are also used because they could be advantageous for certain data or analyses~\cite{paulson:cumulative_sum_scaling,
fisher:habitat_fluctuations,egozcue:isometric_scaling}. 
Any normalization eliminates one dimension of the data and thereby creates compositional biases that complicate the interpretation of the results~\cite{friedman:sparcc, aitchison:original, pawlowsky:book}. For example, the \textit{relative} abundance of a microbe could change simply due to the change in the abundance of other members in the community; such a possibility makes it difficult to unambiguously determine whether this microbe is associated with host phenotype. While it is impossible to fully eliminate compositional biases, their effects could be minimized. In this section, we show that the procedure that we adopted to compute~$C^{-1}$ achieves such minimization for a particular choice of the normalization scheme. We also discuss how DAA can be generalized for an arbitrary normalization scheme and show that the same results are obtained with and without the normalization of the data prior to the analysis. Overall, we conclude that compositional biases do not affect the performance of DAA for diverse microbial communities such as the gut and sample size less than about~$5000$. The application of DAA to data with strong compositional effects would require the modifications that we outline below.

In this section, we use~$l_{i}$ to denote the log-transformed abundance of microbe~$i$ regardless of the normalization scheme. The log-transformation is an important step in the analysis of compositional data because it reduces the degree of compositional biases~\cite{friedman:sparcc, aitchison:original, pawlowsky:book,paulson:cumulative_sum_scaling, fisher:habitat_fluctuations, egozcue:isometric_scaling}. Any normalization of the data imposes a constraint on~$l_i$, which can be stated as follows

\begin{equation}
F(\{l_i\})=0
\label{eq:constraint_F}
\end{equation}

The normalization that we used so far, known as total-sum scaling~\cite{paulson:cumulative_sum_scaling}, corresponds to

\begin{equation}
F(\{l_i\})=-1+\sum_{i}e^{l_i}
\label{eq:total_sum}
\end{equation}

while another popular normalization scheme, known as centered-log ratio, corresponds to

\begin{equation}
F(\{l_i\})=\sum_{i}l_i
\label{eq:centered_log}
\end{equation}

The requirement that~$F(\{l_i\})=0$ changes the maximum entropy distribution to

\begin{equation}
P(\{l_{i}\}) = \delta(F(\{l_i\})) \frac{1}{Z_F} e^{\sum_{i}h_{i}l_{i} + \frac{1}{2}\sum_{ij}J_{ij}l_{i}l_{j}}
\label{eq:maxent_F_hard}
\end{equation}

where~$\delta(\cdot)$ is the Dirac delta function, and the subscript on~$Z$ indicates that the normalization constant depends on the choice of~$F$. It is easy to show the origin of Eq.~(\ref{eq:maxent_F_hard}) by replacing the hard constraint in Eq.~(\ref{eq:constraint_F}) by a soft constraint on the moments of~$P(\{l_{i}\})$. Hard constraints are rarely included in the maximum entropy models while the inclusion of soft constraints is the standard practice. Specifically, we can replace Eq.~(\ref{eq:constraint_F}) by

\begin{equation}
\begin{aligned}
& \langle F(\{l_i\}) \rangle =0\\
& \langle F^2(\{l_i\}) \rangle =\theta^2\\
\end{aligned}
\label{eq:soft_constraint}
\end{equation}

which is equivalent to Eq.~(\ref{eq:constraint_F}) in the limit of~$\theta\to0$. The maximum entropy distribution satisfying Eq.~(\ref{eq:soft_constraint}) reads

\begin{equation}
P(\{l_{i}\}) = \frac{1}{Z_\theta} e^{\sum_{i}h_{i}l_{i} + \frac{1}{2}\sum_{ij}J_{ij}l_{i}l_{j}}e^{-\frac{F^2(\{l_i\}) }{2\theta^2}}
\label{eq:maxent_F_soft}
\end{equation}

which reduces to Eq.~(\ref{eq:maxent_F_hard}) as~$\theta\to0$.

The delta function or the new~$\theta-$dependent term changes the maximum entropy distribution, and Eq.~(\ref{eq:mC_computed}) no longer hold for a general choice of~$F(\{l_i\})$. Instead, one has to compute the first and second order moment of the distribution given by Eq.~(\ref{eq:maxent_F_hard}) or Eq.~(\ref{eq:maxent_F_soft}) and fit them to the means and covariances observed in the data. This procedure, however, cannot uniquely determine~$h_i$ and~$J_{ij}$ because these parameters are no longer independent. Indeed, the condition that~$\langle F^2(\{l_i\}) \rangle=0$ imposes a constraint on the values that~$h_i$ and~$J_{ij}$ can take. This constraint is the consequence of the fact that normalization destroys one dimension of the data. The maximum entropy model ``inherits'' this property, so any change in~$h_i$ could in part be due to the compositional bias.

Accounting for compositional affects for an arbitrary~$F$ is nontrivial and is hardly justified given the weak compositional effects in the IBD data set. The analysis is, however, quite straightforward for~$F$ given by Eq.~(\ref{eq:centered_log}), which corresponds to the normalization by the geometric rather than arithmetic mean of microbial abundances. We now use this choice of~$F$ to illustrate the general principles outlined above and to demonstrate that our implementation of DAA already accounts for the compositional bias for this normalization scheme. 

For~$F$ given by Eq.~(\ref{eq:centered_log}), the soft constraint introduces a factor that keeps~$P(\{l_{i}\})$ a multivariate Gaussian distribution. Therefore, Eq.~(\ref{eq:maxent_F_soft}) is equivalent to our original model given by Eq.~(\ref{eq:maxent_SI}) with~$J$ replaced by~$J^{(\theta)}$ defined as

\begin{equation}
J^{(\theta)}_{ij}= - \frac{1}{\theta^2} + J_{ij}
\end{equation}

In the matrix notation, this definition takes the following form

\begin{equation}
J^{(\theta)}= - \frac{1}{\theta^2}E + J
\end{equation}

\noindent where~$E$ is the matrix with all elements equal to~$1$.

Equations~(\ref{eq:mC_computed}) then continue to hold and can be used to infer~$h^{(\theta)}$ and~$J^{(\theta)}$. As~$\theta\to0$,~$J^{(\theta)}\to J$ in the subspace of~$\sum_{i}l_{i}=0$, i.e. except in the direction of~$(1,1,...,1,1)^{T}$, which becomes the eigenvector of~$J^{(\theta)}$ with a very large eigenvalue. This direction is also an eigenvector of~$C$, and the corresponding eigenvalue tends to zero. Thus, compositional effects render~$C$ degenerate. Strong microbial interactions can have the same effect, and we indeed found a few vanishingly small eigenvalues of~$C$. The variation in the data along the degenerate directions is eliminated when we calculate~$C^{-1}$ using the singular value decomposition~\cite{stewart:early_svd} as explained in the corresponding section above.

This procedure does not artificially exclude taxa from the analysis. For example, if two microbes are perfectly correlated with each other, DAA reports both as significant associations if their abundances vary between health and disease. Since DAA dramatically reduces the number of associations compared to conventional MWAS, we conclude that most of the spurious associations are driven by microbial interactions rather than the compositional bias. Further, the small number of associations found by DAA with quite different relative abundances makes it unlikely that they arise due to compositional effects.

Nevertheless, the maximum entropy model does ``inherit'' a constraint on the parameters from the compositional nature of the data. For~$F(\{l_{i}\})=\sum_{i}l_i$, it is easy to see that~$\sum_{i}h_i$ cannot be uniquely determined from the data. Indeed, adding the same constant to every~$h_i$ changes the exponent in the expression for~$P(\{l_{i}\})$ by a factor proportional to~$\sum_{i}l_i$, which must vanish due to the delta function. One can then choose an arbitrary value for~$\sum_{i}h_i$, say set it to zero. This condition reflects the residual compositional bias left in the maximum entropy model. Similarly, due to the compositional constraint on~$l_i$, the constraint on~$h_i$ can force~$h_i$ to be different for all taxa, even if only one of them is directly affected by the host phenotype. The effect of the constraint, however, should scale as one over the number of the taxa that fluctuate independently. For a diverse ecosystem such as the gut, the effect of the compositional bias should, therefore, be small and detectable only with very large sample sizes. In the synthetic data, we start seeing the compositional effects at about~$5000$ samples which is~$10$ times the number of samples in the IBD data set; see Fig.~\ref{fig:alternative_synthetic_effects1}.

To test for compositional biases in the results of DAA, we analyzed the IBD data set with several widely-used normalization schemes [55, 59], including total-sum scaling, centered-log ratio, cumulative sum scaling, and no normalization at all (Figs. S10 and S13). All analyses  identified about the same number of associations (and the same taxa) using either traditional MWAS or DAA.\\
Finally, we note that our synthetic data has the same amount of compositional bias as in the IBD data. For both data sets, the top 10 most abundant taxa account for 80 \% of the reads, and  we normalized the synthetic data by the total number of reads in the sample prior to performing DAA. 



\textbf{\large{Generation of synthetic data} }

Here, we describe how we generated the synthetic data shown in Fig.~2A of the main text. This data was generated to evaluate the likelihood of spurious associations in MWAS. We introduced a known number of direct associations, but ensured that all other properties of the data correspond to that of the human gut microbiota.

The data for the control group were directly subsampled from the IBD data set. To generate the data for the disease group, we first inferred the covariance matrix using the entire data set and the mean abundances using just the control group.  Then, equation~(\ref{eq:mC_computed}) was used to compute~$h$. These values of~$h$ described normal microbial abundances in subjects without IBD. To introduce a difference between cases and controls, we modified the values of~$h$ for 6 randomly chosen species by~10\% - 40\%; these are typical changes in~$h$ identified by DAA. Finally, we computed the expected microbial abundance using equation~(\ref{eq:mC_computed}) and then sampled from a multivariate Gaussian distribution with these means and the covariance matrix defined above.

We also tested that our conclusions hold for other diseases with potentially different effect sizes. Specifically, we repeated the analysis in Fig.~2A for two other synthetic data sets: one with smaller and one with larger effect sizes. The results are qualitatively similar to what we reported in the main text and are shown in Fig.~\ref{fig:alternative_synthetic_effects1}. The values of the effect sizes are given in Tab.~\ref{tab:synthetic_data}.

\textbf{\large{Data analysis} }

For correlation analysis, we used Pearson correlation coefficient for log-transformed abundances.
 
For logistic regression classifier, we used L1 penalty to ensure sparseness and generalizability. In all classifiers default parameters were used in scikit-learn version 0.17.2. 

For hierarchical clustering of the correlation matrix, we used the Nearest Point Algorithm method of the linkage function in scipy with a correlation distance metric.

\textit{Threshold for matrix inversion}\\
For our analysis of the IBD and synthetic data sets we set~$\lambda_{\mathrm{min}}$ to~$0.01$. To test whether our results are robust to the value of the threshold, we varied the number of eigenvalues of~$\Lambda^{-1}$ not set to zero; see Fig.~\ref{fig:EV_threshold}. When only a few eigenvalues where included, DAA detected a large number of associations because many taxa were perfectly correlated, and it was impossible to distinguish direct from indirect associations. As the number of included eigenvalues increased, the performance of DAA improved and reached a plateau. In this plateau region, the results were largely insensitive to the value of the threshold used. Our choice of the theshold corresponded to this plateau region. At all taxonomic levels, we found one or two almost zero eigenvalues that were below~$\lambda_{min}$~(Fig.~\ref{fig:eigenvalues_uncertainty}); all other eigenvalues were included in the analysis.

\textbf{\large{Computer code} }

We include here the link to computer code that loads the data and outputs all figures and tables: https://github.com/rajitam/DAA-figures-and-tables

\newpage

\begin{figure}
\begin{center}
\includegraphics[width = 6in]{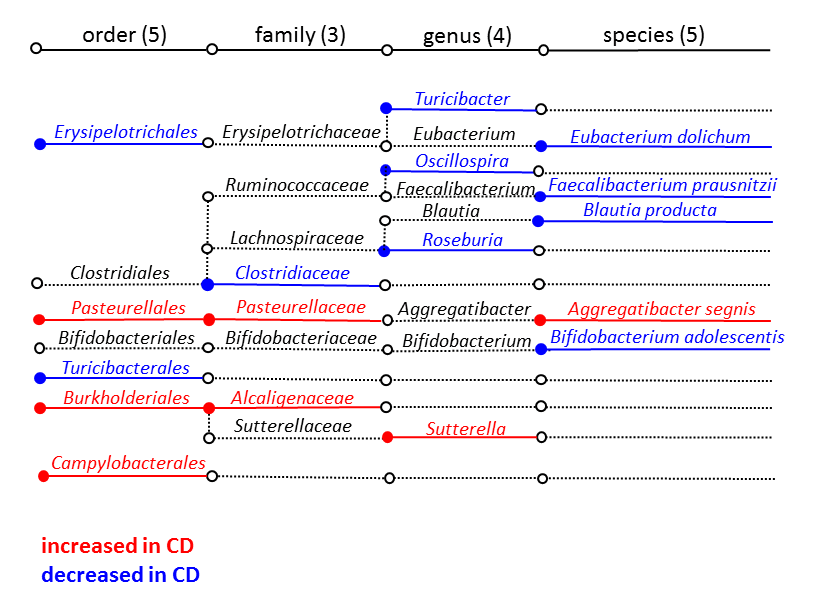}
\caption{\textbf{Taxa directly associated with Crohn's disease.} Note that the Green Genes database~\cite{green:genes} used in QIIME~\cite{caporaso:qiime} places Turicibacter under Erysipelotrichales and has a unique order of Turicibacterales. This apparent inconsistency may reflect insufficient understanding of Turicibacter phylogeny. The effect sizes and statistical significance are summarised in Tab.~\protect{\ref{tab:daa_results}} and compared between DAA and conventional MWAS in Tab.~\protect{\ref{tab:hl_comparison}}.}
\label{fig:direct_associations}
\end{center}  
\end{figure}

\begin{figure}
\begin{center}
\includegraphics[width = 6in]{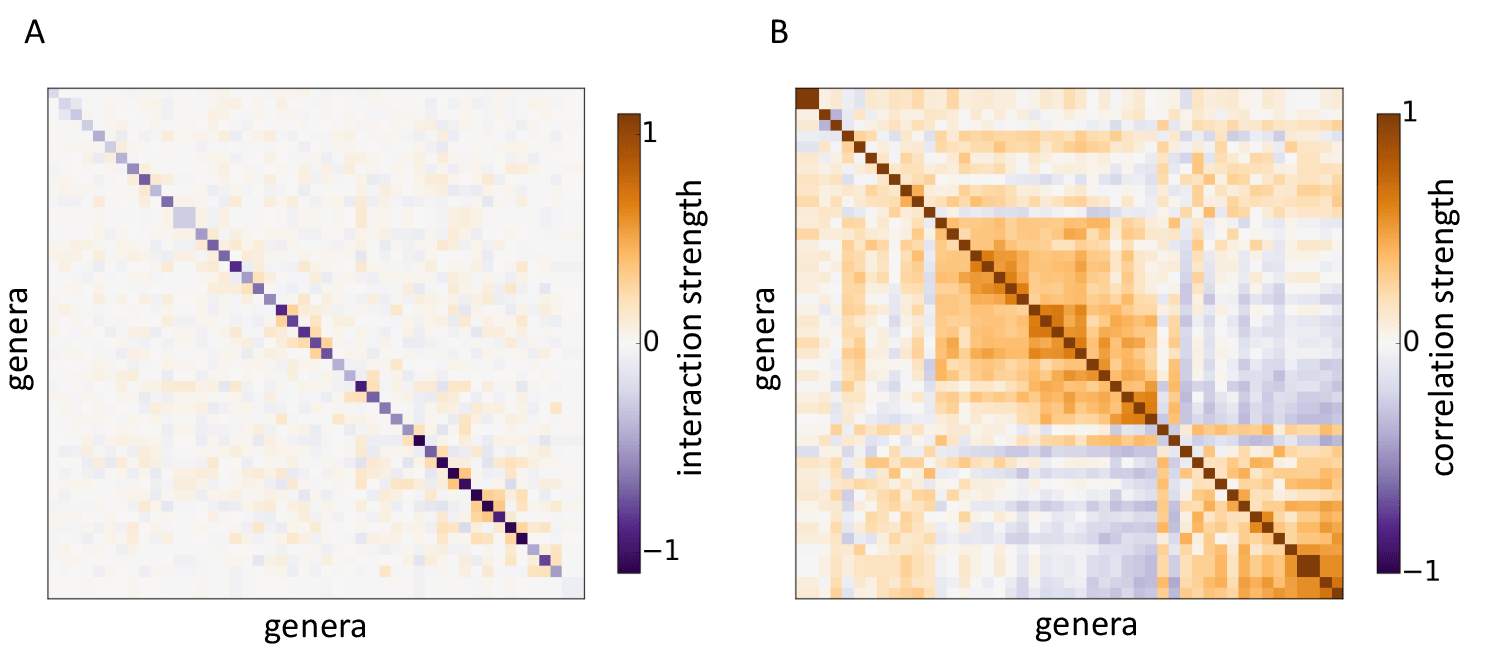}
\caption{\textbf{Comparison between correlations and direct interactions.} The matrix of microbial interactions~$J$ is shown in~\textbf{(A)} and the correlation matrix~$C$ is shown in~\textbf{(B)}, which is the same as Fig.~1B of the main text. Both matrices are inferred from the IBD data set. Note that~$J$ is sparser than~$C$. For greater clarity, the matrices are hierarchically clustered; therefore, the order of species in~A and~B is not the same.}
\label{fig:interaction_matrix}
\end{center}  
\end{figure}

\begin{figure}
\begin{center}
\includegraphics[width = 6.5in]{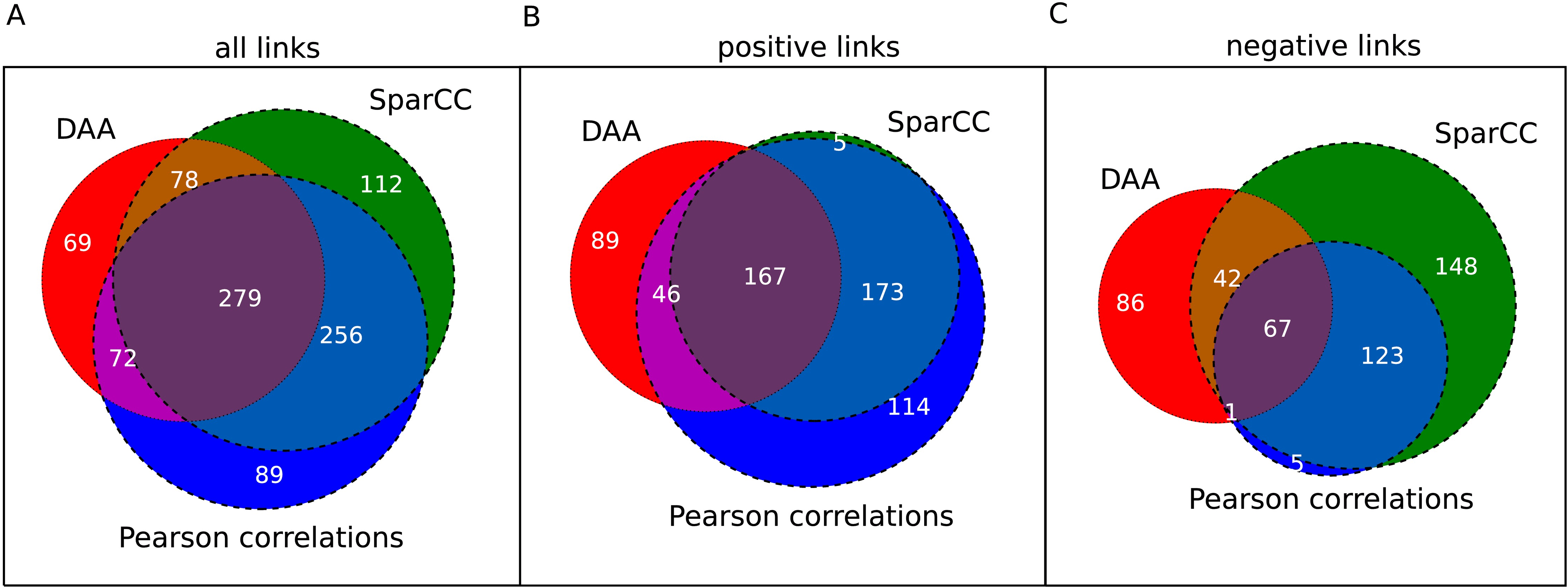}
\vspace{.15in}
\caption{\textbf{Comparison of networks inferred by Pearson correlation, SparCC, and DAA at the genus level.} Three networks quantifying microbial co-occurrence or interactions have been inferred: one based on the Pearson correlation coefficient between log-abundances~(which is closely related to the covariance matrix~$C$), one using SparCC package from Ref.~\cite{friedman:sparcc} that attempts to reduce compositional bias, and one based on the direct interactions~$J$ from DAA. In each network, we kept only links that were statistically different from~$0$ under a permutation test with 5\% false discovery rate. The panels display Venn diagrams showing unique and overlapping links in these networks. All links are included in \textbf{(A)}, and the comparison is done irrespective of the sign of the link, i.e. agreement is reported even if one method reports a positive link and another method reports a negative link. In contrast, \textbf{(B)}~and~\textbf{(C)} show only positive and negative links respectively. Three conclusions can be drawn from these comparisons. First, the high overlap between SparCC and Pearson networks shows that log-transforms have largely accounted for the compositional bias. Second, all three methods agree on a large number of links suggesting that all methods are sensitive to some strong interactions. Third, DAA reports fewer links and identifies a few links not detected by other methods. This reflect the different nature of DAA links. While both Pearson correlation and SparCC infer correlation, which could be either direct or indirect~(i.e. induced; see main text). DAA removes indirect correlations, thus reducing the total number of links, but also reveals pairwise interactions that could have been masked by strong correlations with a third species.}
\label{fig:interactions_comparison}
\end{center}  
\end{figure}

\begin{figure}
\begin{center}
\includegraphics[width = 6in]{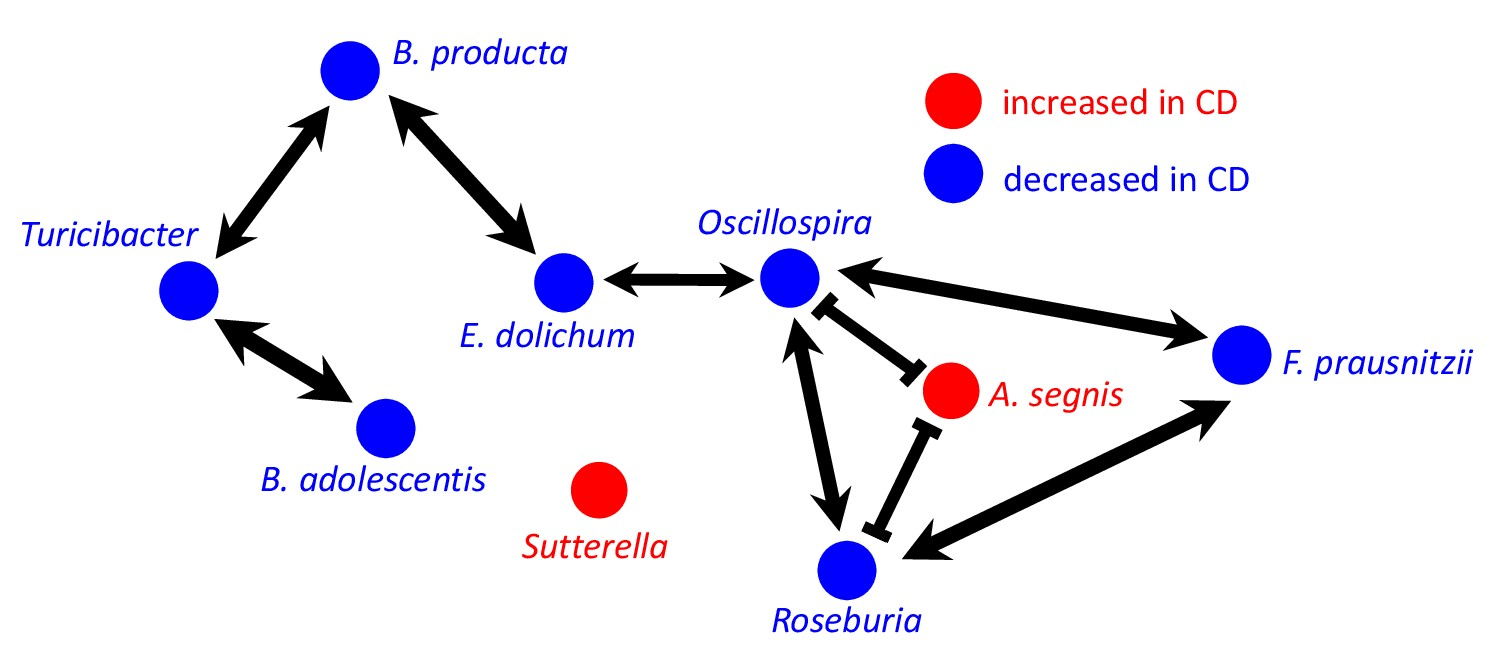}
\vspace{.15in}

\caption{\textbf{The network based on the correlation coefficient between log-transformed abundances.} We plotted the correlation-based network for the species detected by DAA. Note the similarities and differences with the interaction network shown in Fig.~3 of the main text. Only the links with the correlation coefficient greater than 0.27 or lower than -0.15 are shown, and all links are statistically significant~($q<0.05$). All correlation coefficients and direct interactions are summarized in Tab.~\protect{\ref{tab:interactions}} for the genera and species detected by DAA.}
\label{fig:correlation_matrix}
\end{center}  
\end{figure}

\begin{figure}
\begin{center}
\includegraphics[width = 6in]{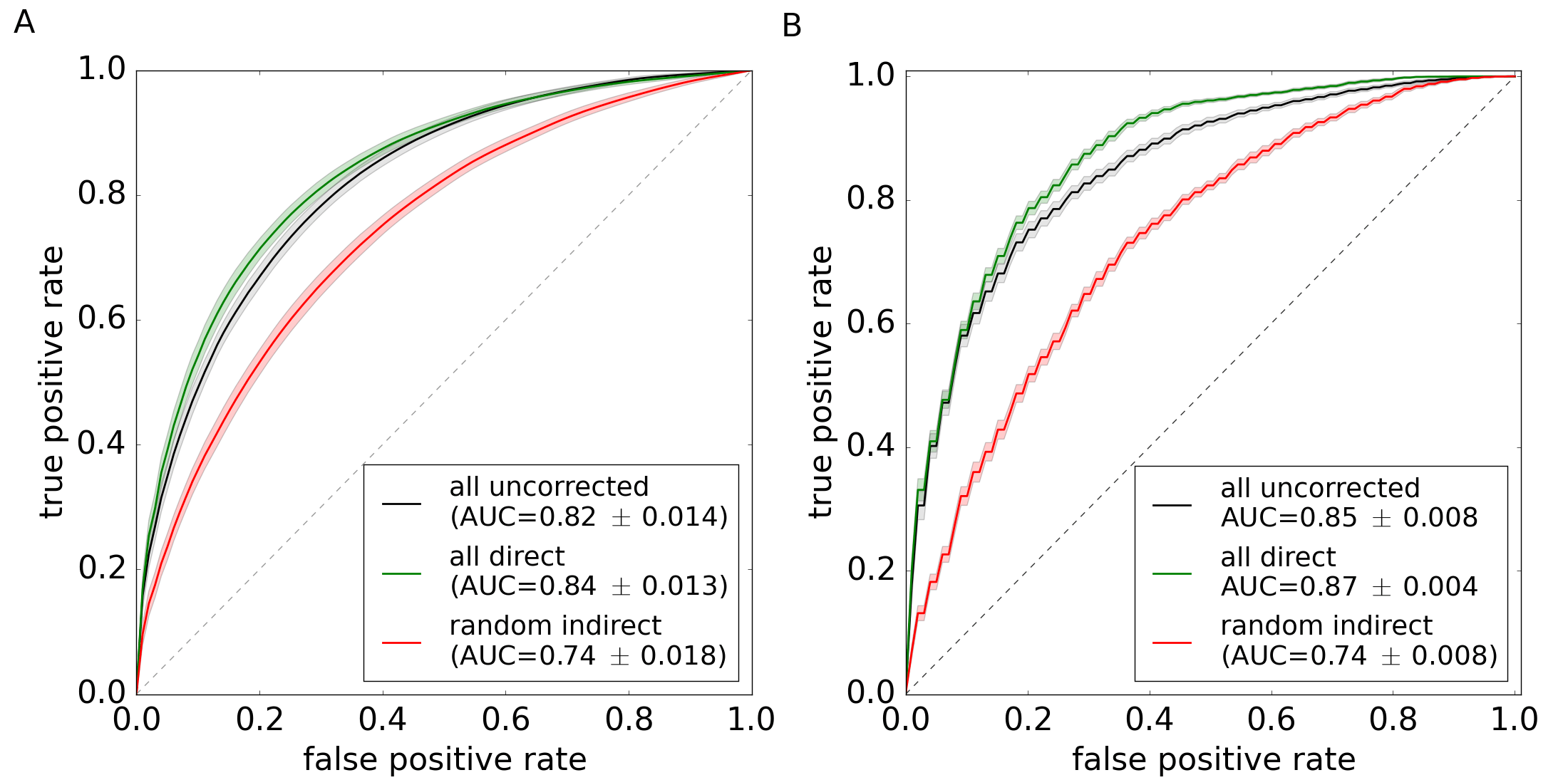}
\caption{\textbf{Direct associations retain full diagnostic power.} The same as Fig.~4B of the main text, but for two other classifiers: random forest~\cite{ho:random_forest, breiman:random_forest} in~\textbf{(A)} and support vector machine~\cite{cortes:svm} in~\textbf{(B)}.}
\label{fig:other_classifiers}
\end{center}  
\end{figure}

\begin{figure}
\begin{center}
\includegraphics[width = 3.7in]{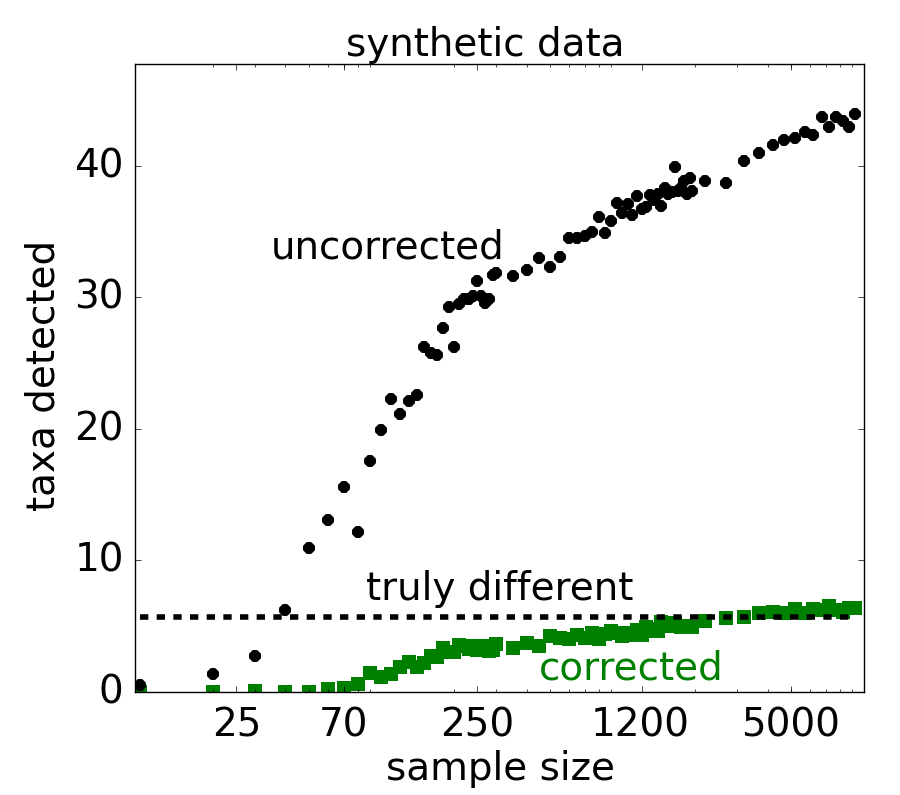}

\caption{\textbf{DAA detects all directly associated taxa in synthetic data, provided the sample size is sufficiently large.} The same as Fig.~2A in the main text, but with the $x$-axis extended to larger sample sizes. Note that DAA recovers all 6 directly associated taxa when the sample size is greater than about 1200.}
\label{fig:Figure2_full}
\end{center}  
\end{figure}

\begin{figure}
\begin{center}
\includegraphics[width = 6in]{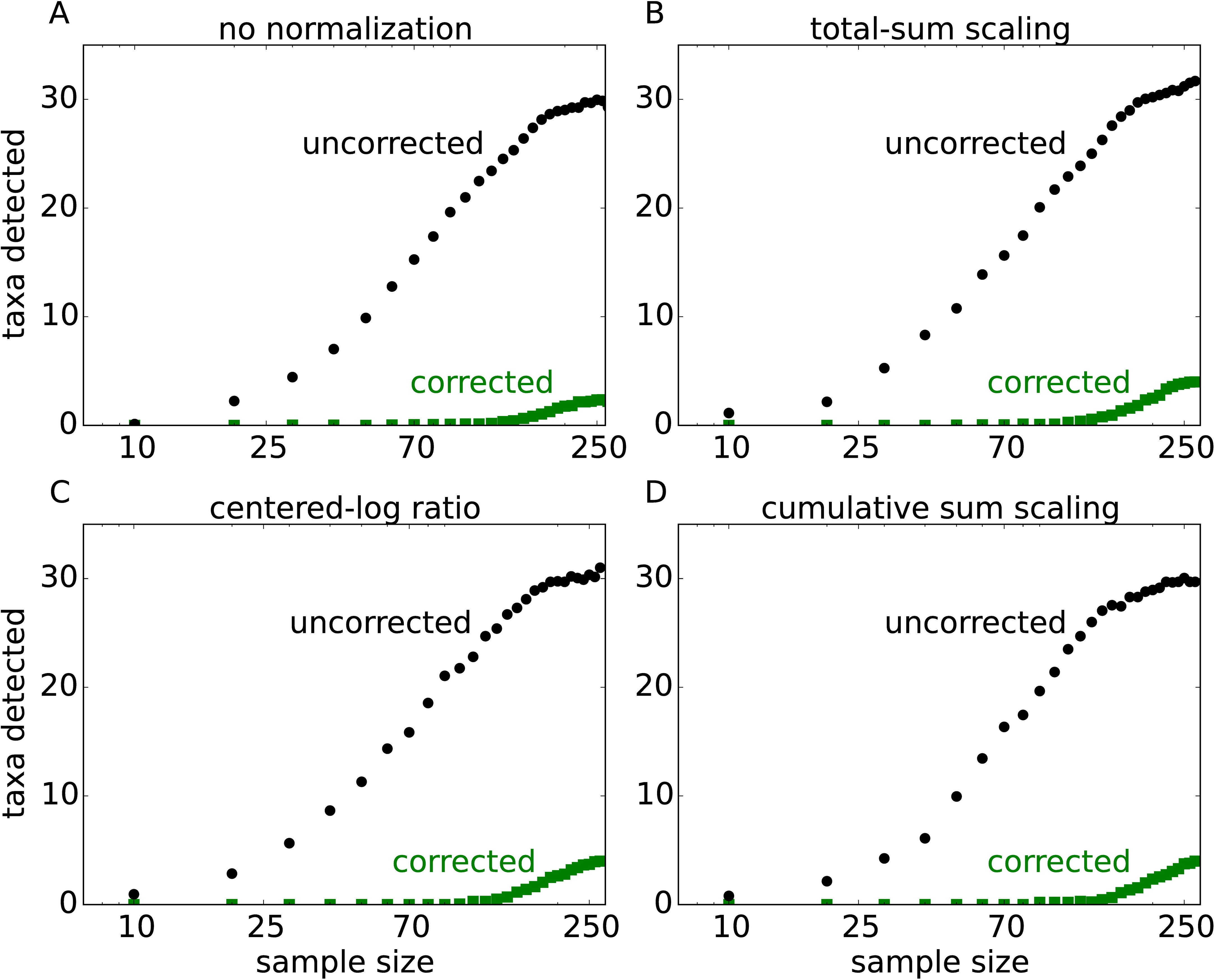}

\caption{\textbf{Compositional bias has a negligible effect on DAA performance.} All panels are the same as Fig.~2C in the main text, but with different normalization of the data prior to the analysis. \textbf{(A)}~No normalization: the analysis is done on the counts from the OTU table, which do not add up to a constant number. \textbf{(B)}~Total-sum scaling: The counts are converted into relative abundances by dividing by the total number of counts~(reads) per sample. This plot is the same as Fig.~2C. \textbf{(C)} Centered-log ratio: First log-abundances were computed from unnormalized counts with a pseudocount of 1. Then, the mean log-abundances of the taxa was computed by averaging over the samples. Finally, the mean-log abundance of every taxon was subtracted from the log-abundances of this taxon in all samples. This procedure corresponds to normalizing by the geometric mean of the counts because it ensures that the mean log-abundance of a taxon is zero [55]. \textbf{(D)} Cumulative sum scaling: A normalization scheme proposed specifically for microbiome analyses was implemented following  Ref.~[59]. The results of the analyses in A-D are very similar suggesting that compositional bias does not lead to major artifacts. In particular, the number of associations in A grows at the same rate with the sample size as in B-D. This would not be the case if the compositional bias was strong because spurious associations due to normalization would lead to a greater number of detected taxa. Thus, we conclude that interspecific interactions rather than compositional effects are the primary source of spurious associations.}
\label{fig:norms_counts}
\end{center}  
\end{figure}

\begin{figure}
\begin{center}
\includegraphics[width = 3.7in]{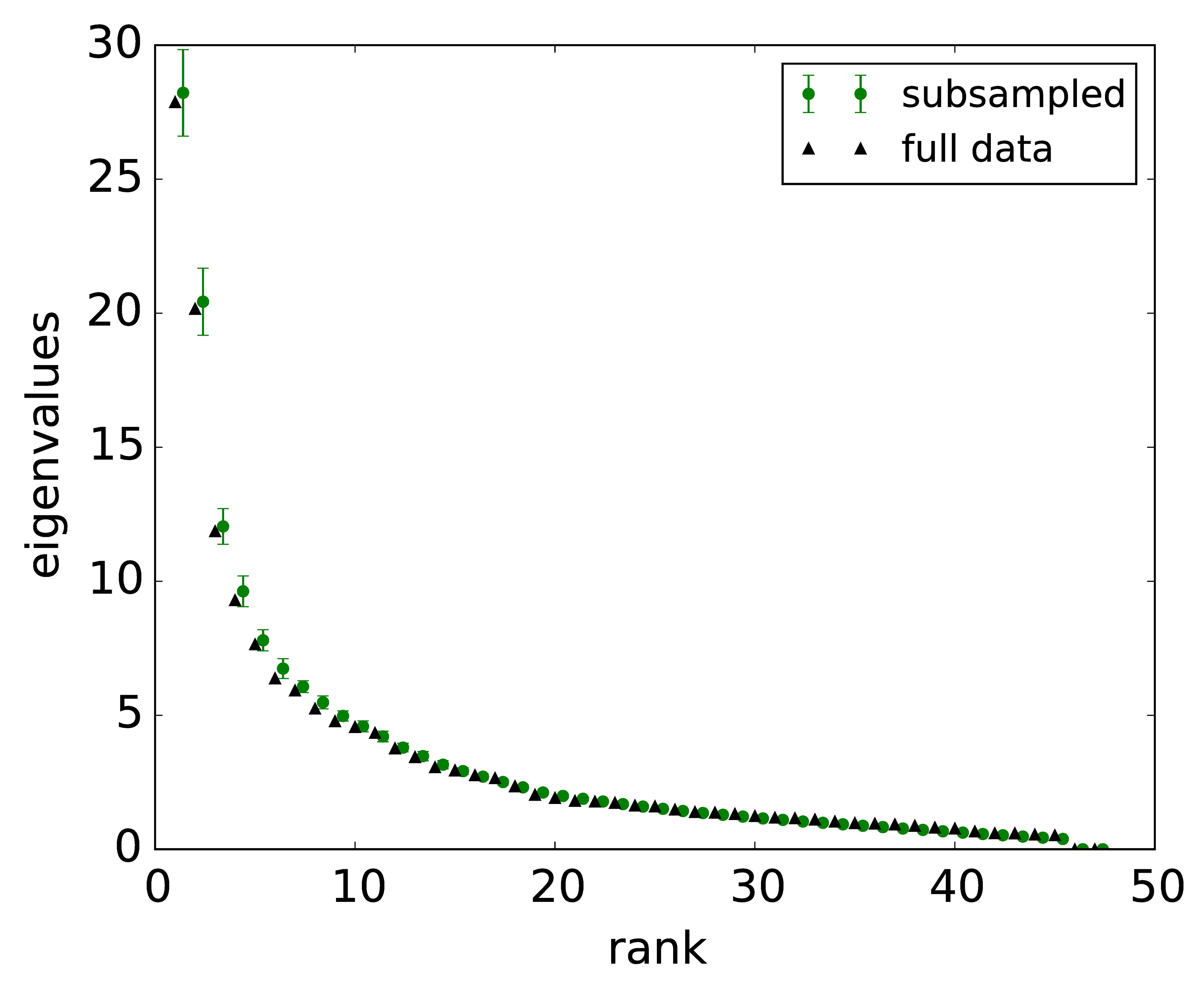}

\caption{\textbf{The inference of the eigenvalues of the covariance matrix is robust to variation in sample size and bootstrapping.} We repeatedly subsampled the IBD data set to half of its size and computed the eigenvalues of the covariance matrix~$C$. The means and standard deviations from this bootstrap procedure are shown in green, and the eigenvalue inferred from the entire data are shown in black. The agreement between the different sample sizes and the small variation due to subsampling indicate that the spectral properties of~$C$ can be inferred quite accurately.}
\label{fig:eigenvalues_uncertainty}
\end{center}  
\end{figure}

\begin{figure}
\begin{center}
\includegraphics[width = 6in]{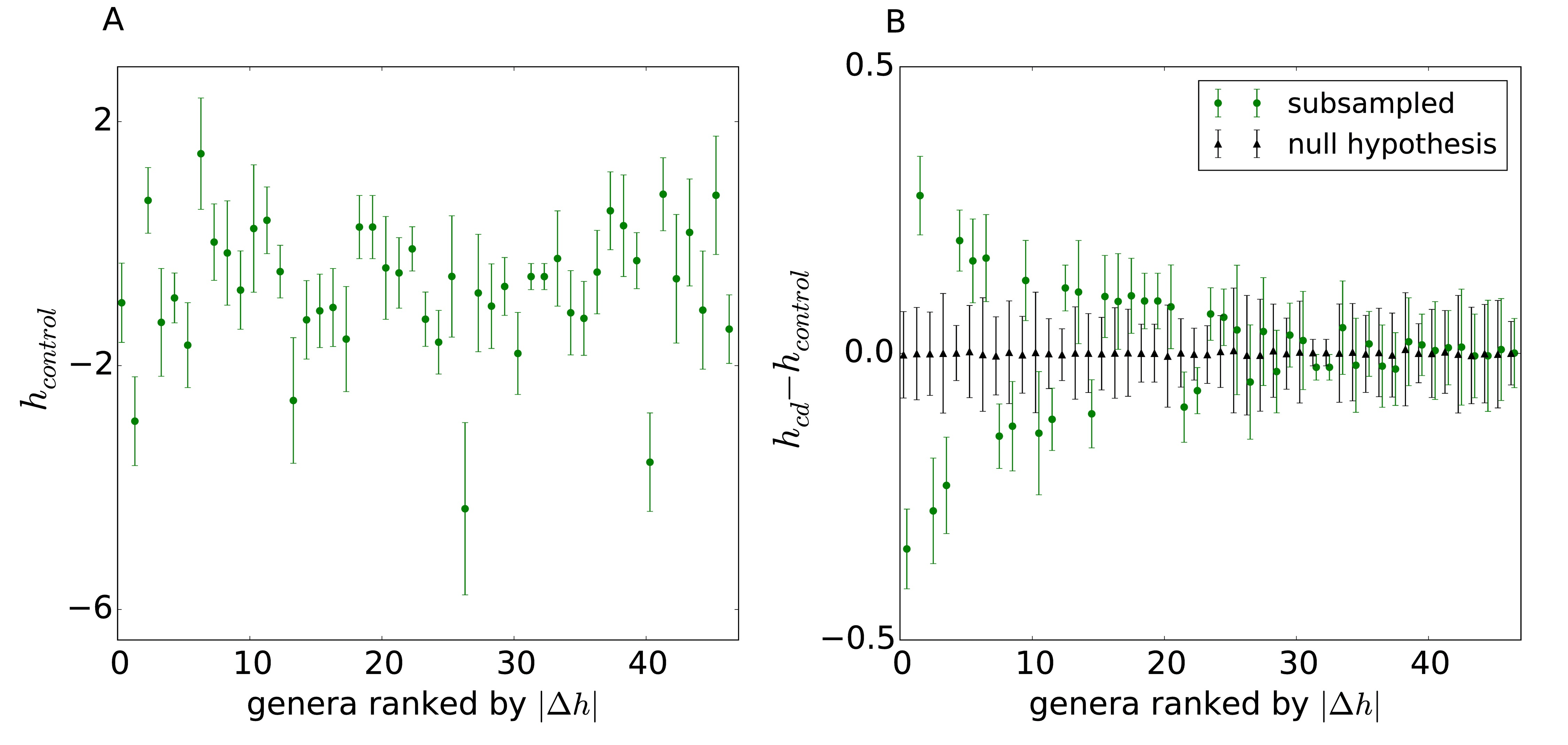}

\caption{\textbf{Results of DAA are robust to variation in sample size and bootstrapping.}  Similar to Fig.~\ref{fig:eigenvalues_uncertainty}, we repeatedly subsampled the IBD data set to half of its size and carried out DAA on each of the subsamples. \textbf{(A)} shows that there is a modest variation in inferred $h$. To a large extent, this variation is driven by the uncertainty in~$C$ and its inverse~$J$. \textbf{(B)} shows a much smaller variation in~$\Delta h$ between control and CD groups~(green symbols). The noise is reduced because, even though~$C$ changes from subsample to subsample, the same~$C$ is used to infer~$h$ for control and disease groups. Therefore, the variability in~$C$ has a much weaker effect on~$\Delta h$. For comparison, we also show~$\Delta h$ obtained by bootstrapping the entire data set without preserving the diagnosis labels~(black symbols). These data show the expected distribution of~$\Delta h$ under the null hypothesis of no associations. For genera detected by DAA, the black and the green error bars do not overlap suggesting that the results of DAA are not affected by the uncertainty in~$C$ and are robust to variation in sample size and bootstrapping.}
\label{fig:dh_subsampled}
\end{center}  
\end{figure}

\begin{figure}
\begin{center}

\includegraphics[width = 3.7in]{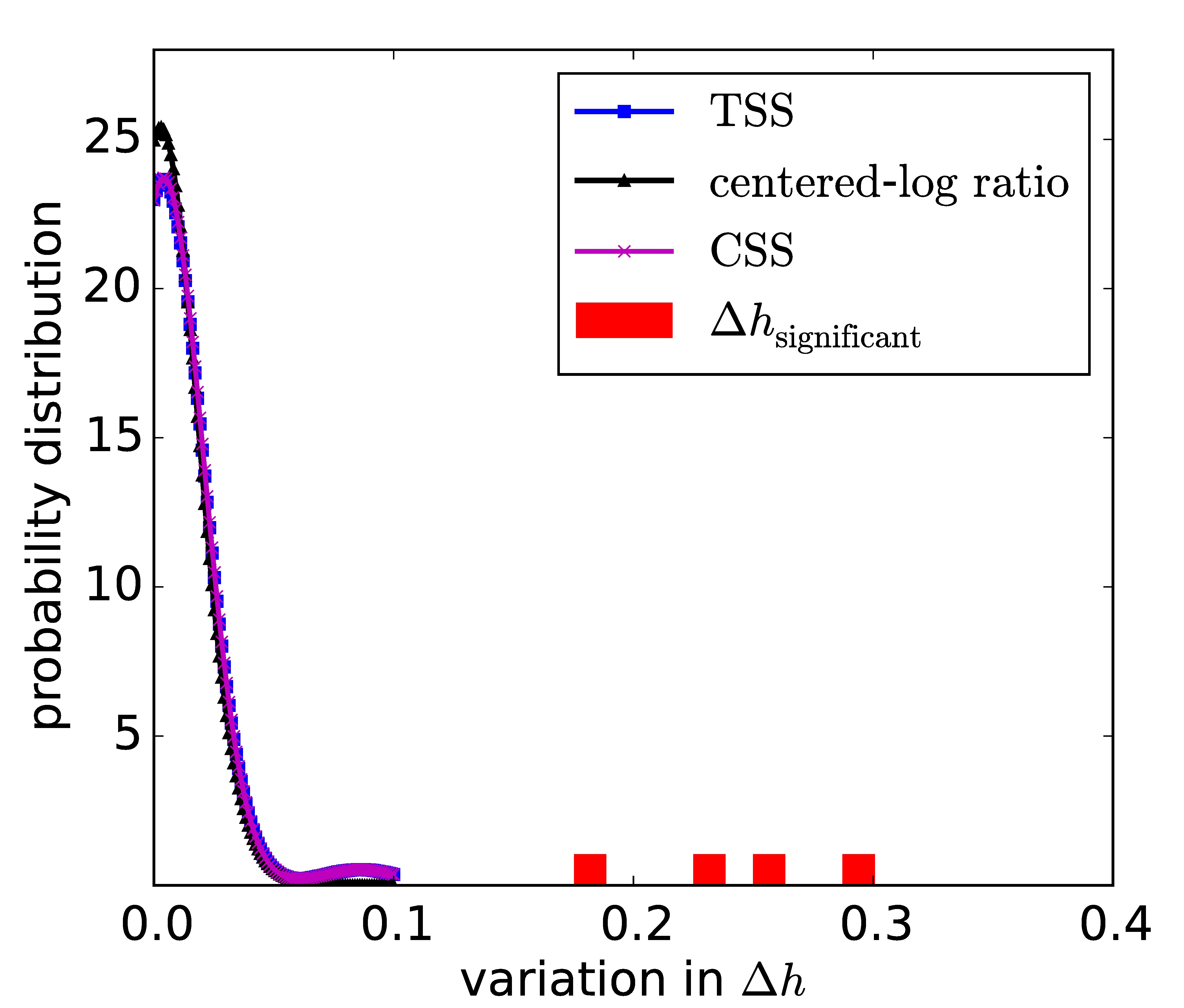}
\vspace{.15in}
\caption{\textbf{Results of DAA are not significantly affected by compositional effects.} The quantity $\Delta h$ between control and CD groups is the test statistic used to infer direct associations, and the variation of~$\Delta h$ due to sampling shows whether the statistical analysis is robust to small changes in the data set. To quantify these variations in $\Delta h$, we consider a sample drawn from the maximum entropy model fitted to the IBD data set and define two \textbf{$\delta\Delta h$}: one between normalized and not normalized sample and the other between the not normalized sample and the values of~$h$ in the maximum entropy model. The first $\delta\Delta h$ quantifies the variability due to normalization, while the second $\delta\Delta h$ quantifies the variability due to sampling. The plot shows the distribution of the absolute values of the difference between the absolute values of these $\delta\Delta h$ across genera for three normalization schemes: total-sum scaling (TSS), centered-log ratio (CLR) and cumulative sum scaling (CSS). The absolute $\Delta h$ values of significant taxa in IBD RISK data (red rectangles) lie well outside of the distributions shown. }
\label{fig:compbias_dist}
\end{center}  
\end{figure}

\begin{figure}
\begin{center}
\includegraphics[width = 6in]{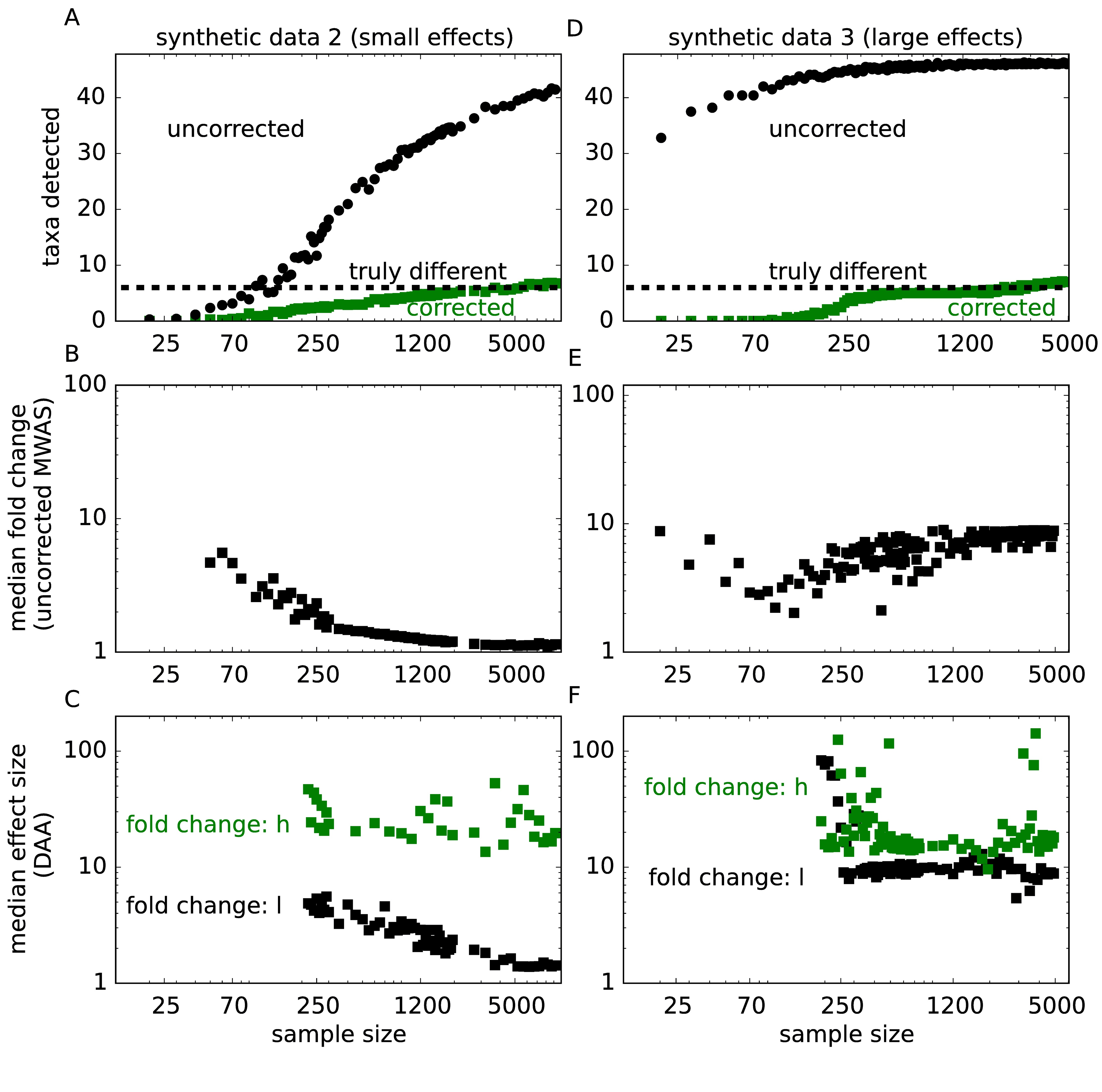}

\caption{\textbf{Spurious associations in synthetic data with small and large effect sizes.} The same analysis as in Fig.~2AB of the main text, but for synthetic data with smaller~(A, B, C) and larger~(D, E, F) effect sizes. \textbf{(A)}~and~\textbf{(D)} show the number of associations detected by traditional MWAS and DAA. \textbf{(B)}~and~\textbf{(E)} show the median effect sizes~(median fold change) for the taxa detected by conventional MWAS. \textbf{(C)}~and~\textbf{(E)} show the effect sizes in both ~h and ~l for the taxa detected by DAA. The effect size for h was quantified as the relative percent difference in host-field between cases and controls, while the l-effect size was computed as described in the main text. Overall the results are similar to those in Fig. 2. In addition, (A)~and~(B) show that DAA can recover all directly associated taxa given a large number of samples without any false positives. For sample sizes exceeding~$5000$, DAA starts to detect indirect associations due to compositional effects.}
\label{fig:alternative_synthetic_effects1}
\end{center}  
\end{figure}
 
\begin{figure}
\begin{center}
\includegraphics[width = 6in]{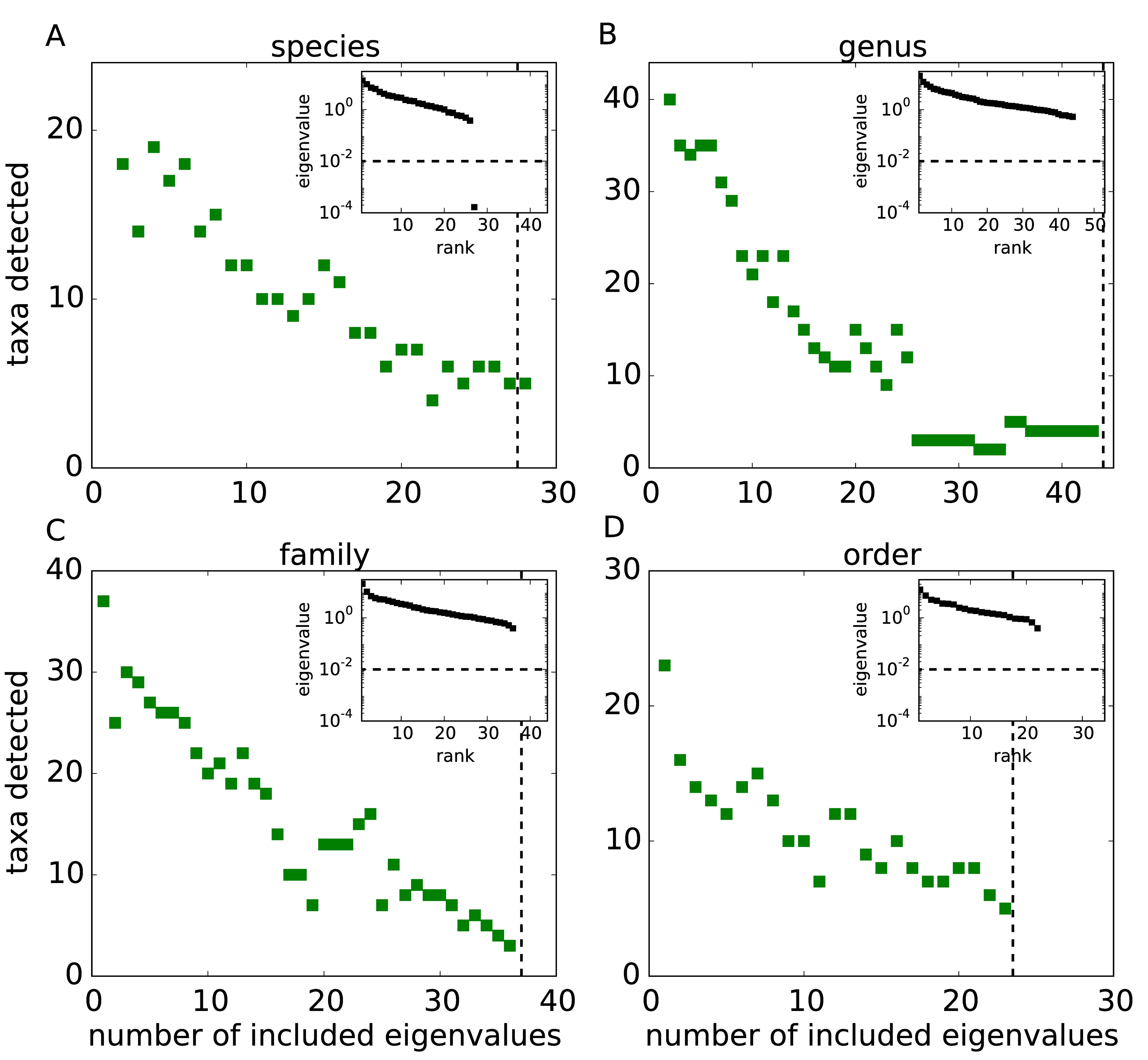}
\vspace{.15in}
\caption{\textbf{ Sensitivity of DAA to eigenvalue threshold~$ \bm{\lambda_{\mathrm{min}}} $.} Large~$\lambda_{\mathrm{min}}$ retains only a few eigenvalues and imposes an artificially strong correlation structure on the data. As a result, DAA detects a large number of associations because it cannot distinguish direct from indirect effects. The performance of DAA improves as more eigenvalues are included and reaches a plateau. The dashed lines show the number of eigenvalues included for~$\lambda_{\mathrm{min}}=0.01$ used throughout our analysis. The insets show the eigenvalues of~$\Lambda$ in decreasing order. The four panels show the results for different taxonomic levels: from species to order.}
\label{fig:EV_threshold}
\end{center}  
\end{figure}

\clearpage
\setlength{\abovecaptionskip}{+5ex}

\begin{table}[]
\centering
\caption{\textbf{The list of genera used in the analysis.} We included all genera that were present in more than 60\% of either control or IBD subjects. The indices were chosen to hierarchically cluster the correlation matrix shown in Fig.~1b of the main text~(index corresponds to the position of the genus on the x axis).}

\label{tab:genera_analyzed}
\begin{tabular}{@{}llllll@{}}
\hline\hline
index & genus name & index & genus name & index & genus name \\ \hline \hline

1 & \textit{{[}Prevotella{]}} & 17 & \textit{Corynebacterium} & 33 & \textit{Fusobacterium} \\
2 & \textit{Prevotella} & 18 & \textit{Pseudomonas} & 34 & \textit{Bacteroides} \\
3 & \textit{Dialister} & 19 & \textit{Acinetobacter} & 35 & \textit{Anaerostipes} \\
4 & \textit{Phascolarctobacterium} & 20 & \textit{Erwinia} & 36 & \textit{Parabacteroides} \\
5 & \textit{Epulopiscium} & 21 & \textit{Actinomyces} & 37 & \textit{{[}Eubacterium{]}} \\
6 & \textit{Eggerthella} & 22 & \textit{Streptococcus} & 38 & \textit{Odoribacter} \\
7 & \textit{Clostridium} & 23 & \textit{Granulicatella} & 39 & \textit{Oscillospira} \\
8 & \textit{Akkermansia} & 24 & \textit{Neisseria} & 40 & \textit{Lachnospira} \\
9 & \textit{Bilophila} & 25 & \textit{Rothia} & 41 & \textit{Roseburia} \\
10 & \textit{Bifidobacterium} & 26 & \textit{Eikenella} & 42 & \textit{Faecalibacterium} \\
11 & \textit{Collinsella} & 27 & \textit{Campylobacter} & 43 & \textit{Dorea} \\
12 & \textit{Sutterella} & 28 & \textit{Veillonella} & 44 & \textit{{[}Ruminococcus{]}} \\
13 & \textit{Parvimonas} & 29 & \textit{Actinobacillus} & 45 & \textit{Ruminococcus} \\
14 & \textit{Porphyromonas} & 30 & \textit{Aggregatibacter} & 46 & \textit{Blautia} \\
15 & \textit{Turicibacter} & 31 & \textit{Haemophilus} & 47 & \textit{Coprococcus} \\
16 & \textit{Staphylococcus} & 32 & \textit{Holdemania} &  & \\ \hline\hline
\end{tabular}
\end{table}

\begin{table}[]
\centering
\caption{\textbf{Genera modified in synthetic data.} Taxa indices are the same as in Table~\protect{\ref{tab:genera_analyzed}}. Effect size is the percent change in the value of~$h$.}
\label{tab:synthetic_data}
\begin{tabular}{@{}llll@{}}
\hline\hline
\textbf{\begin{tabular}[c]{@{}l@{}}taxon\\ index\end{tabular}} & \textbf{\begin{tabular}[c]{@{}l@{}} effect size\\ data 1 (main text) \end{tabular}} & \textbf{\begin{tabular}[c]{@{}l@{}}effect size\\ data 2 (small) \end{tabular}} & \textbf{\begin{tabular}[c]{@{}l@{}}effect size\\ data 3 (large)\end{tabular}} \\ \hline\hline
1 & $-$18\% & $-$17\% & $-$44\% \\
11 & +24\%& +14\%& +129\% \\
19 & $-$36\%& $-$12\%& $-$72\% \\
27 & +17\%& +16\%& +67\% \\
33 & $-$13\%& $-$14\%& $-$28\% \\
45 & +18\%& +13\%& +112\% \\ \hline\hline
\end{tabular}
\end{table}

\clearpage

\begin{table}[]
\centering
\caption{\textbf{Direct associations identified by DAA across phylogenetic levels.}}
\label{tab:daa_results}
\begin{tabular}{@{}llllll@{}}
\hline\hline

\textbf{\begin{tabular}[c]{@{}l@{}}taxon\\ name\end{tabular}} & \textbf{\begin{tabular}[c]{@{}l@{}}direct effect, \\ $\bm{h_{\mathrm{CD}}}$\end{tabular}} & \textbf{\begin{tabular}[c]{@{}l@{}}direct effect, \\ $\bm{h_{\mathrm{ctrl}}}$\end{tabular}} & \textbf{\begin{tabular}[c]{@{}l@{}}difference,\\ $ \bm{\Delta h}/\vert\bm{h_{\mathrm{ctrl}}}\vert$\end{tabular}} & \textbf{p-value} & \textbf{q-value}\\ \hline\hline \\
\multicolumn{6}{c}{\textbf{Order level}} \\
\textit{Burkholderiales} & $-$0.47 & $-$0.66 & +0.29 & 0.00013 & 0.0029 \\
\textit{Turicibacterales} & $-$1.7 & $-$1.4 & $-$0.18 & 0.00031 & 0.0036 \\
\textit{Pasteurellales} & $-$0.51 & $-$0.69 & +0.26 & 0.00068 & 0.0052 \\
\textit{Campylobacterales} & $-$1.6 & $-$1.8 & +0.1 & 0.00696 & 0.04 \\
\textit{Erysipelotrichales} & $-$2.5 & $-$2.3 & $-$0.083 & 0.0095 & 0.044 \\ \\
\multicolumn{6}{c}{\textbf{Family level}} \\
\textit{Alcaligenaceae} & $-$0.68 & $-$0.86 & +0.21 & 0.00027 & 0.01 \\
\textit{Clostridiaceae} & $-$1.2 & $-$0.99 & $-$0.18 & 0.0026 & 0.049 \\
\textit{Pasteurellaceae} & $-$0.31 & $-$0.47 & +0.35 & 0.0033 & 0.049 \\ \\
\multicolumn{6}{c}{\textbf{Genus level}} \\
\textit{Roseburia} & $-$1.2 & $-$0.86 & $-$0.35 & 0.000098 & 0.0046 \\
\textit{Sutterella} & $-$0.63 & $-$0.80 & +0.22 & 0.00043 & 0.01 \\
\textit{Oscillospira} & $-$2.4 & $-$2.6 & +0.097 & 0.0015 & 0.023 \\
\textit{Turicibacter} & +0.46 & +0.69 & $-$0.34 & 0.003 & 0.035 \\ \\
\multicolumn{6}{c}{\textbf{Species level}} \\
\textit{B.adolescentis} & $-$0.23 & +0.073 & $-$4.12 & 0.00013 & 0.0037 \\
\textit{E.dolichum} & $-$0.51 & $-$0.31 & $-$0.65 & 0.0028 & 0.039 \\
\textit{F.prausnitzii} & $-$0.97 & $-$0.81 & $-$0.20 & 0.0042 & 0.039 \\
\textit{A.segnis} & $-$0.072 & $-$0.25 & +0.71 & 0.0056 & 0.04 \\
\textit{B.producta} & $-$0.75 & $-$0.54 & $-$0.38 & 0.0064 & 0.04 \\\\ \hline\hline

\end{tabular}
\end{table}

\clearpage

\begin{table}[]
\centering
\caption{\textbf{Comparison between changes in~$\bm{h}$ and in~$\bm{l}$ for the taxa identified by DAA.} }
\label{tab:hl_comparison}
\begin{tabular}{lllll}
\hline\hline
\textbf{\begin{tabular}[c]{@{}l@{}}taxon \\ name\end{tabular}} & \textbf{\begin{tabular}[c]{@{}l@{}}abundance \\ $\bm{l_{\mathrm{CD}}/l_{\mathrm{ctrl}}}$\end{tabular}} & \textbf{\begin{tabular}[c]{@{}l@{}} direct effect \\  $ \bm{\Delta h}/\vert\bm{h_{\mathrm{ctrl}}}\vert$ \end{tabular}} & \textbf{q-value, $l$} & \textbf{q-value, $h$} \\ \hline\hline
 \\
\multicolumn{5}{c}{\textbf{Order level}} \\
\textit{Burkholderiales} & +1.6 & +0.29 & 0.04 & 0.0029 \\
\textit{Turicibacterales} & +0.45 & $-$0.18 & 0.00002 & 0.0036 \\
\textit{Pasteurellales} & +4.2 & +0.26 & 0 & 0.0052 \\
\textit{Campylobacterales} & +2.1 & +0.1 & 0.000001 & 0.04 \\
\textit{Erysipelotrichales} & +0.34 & $-$0.083 & 0 & 0.044 \\
 &  &  &  &  \\
\multicolumn{5}{c}{\textbf{Family level}} \\
\textit{Alcaligenaceae} & +1.7 & +0.21 & 0.03 & 0.01 \\
\textit{Clostridiaceae} & +0.25 & $-$0.18 & 0 & 0.049 \\
\textit{Pasteurellaceae} & +4.2 & +0.35 & 0 & 0.049 \\
 &  &  &  &  \\
\multicolumn{5}{c}{\textbf{Genus level}} \\
\textit{Roseburia} & +0.21 & $-$0.35 & 0 & 0.0046 \\
\textit{Sutterella} & +2.0 & +0.22 & 0.004 & 0.01 \\
\textit{Oscillospira} & +0.84 & +0.097 & 0.33 & 0.023 \\
\textit{Turicibacter} & +0.50 & $-$0.34 & 0.0004 & 0.035 \\
 &  &  &  &  \\
\multicolumn{5}{c}{\textbf{Species level}} \\
\textit{B.adolescentis} & +0.43 & $-$4.12 & 0.00004 & 0.0037 \\
\textit{E.dolichum} & +0.43 & $-$0.65 & 0.00004 & 0.039 \\
\textit{F.prausnitzii} & +0.41 & $-$0.20 & 0.000003 & 0.039 \\
\textit{A.segnis} & +2.8 & +0.71 & 0 & 0.04 \\
\textit{B.producta} & +0.67 & $-$0.38 & 0.03 & 0.04
\\\\ \hline\hline

\end{tabular}
\end{table}

\clearpage

\begin{table}[]
\small
\centering

\caption{\textbf{Indirect associations identified by uncorrected abundance analysis across phylogenetic levels.}}
\label{tab:indirect_results}

\begin{tabular}{@{}llllll@{}}
\hline \hline 
\textbf{taxon name} & \textbf{\begin{tabular}[c]{@{}l@{}}abundance, \\ $\bm{l_{\mathrm{CD}}}$ \end{tabular}} & \textbf{\begin{tabular}[c]{@{}l@{}}abundance,\\ $\bm{l_{\mathrm{ctrl}}}$\end{tabular}} & \textbf{\begin{tabular}[c]{@{}l@{}}ratio,\\ $\bm{l_{\mathrm{CD}}/l_{\mathrm{ctrl}}}$\end{tabular}} & \textbf{p-value} & \textbf{q-value} \\ \hline \hline \\
\multicolumn{6}{c}{\textbf{Order level}} \\
\textit{Erysipelotrichales} & 0.43 & 1.3 & 0.34 & 0 & 0 \\
\textit{Clostridiales} & 18.4 & 31.1 & 0.59 & 0 & 0 \\
\textit{Pasteurellales} & 1.2 & 0.29 & 4.2 & 0 & 0 \\
\textit{Fusobacteriales} & 0.25 & 0.08 & 3.2 & 0 & 0 \\
\textit{Enterobacteriales} & 2.8 & 0.81 & 3.4 & 0 & 0 \\
\textit{Campylobacterales} & 0.017 & 0.008 & 2.1 & 0.000001 & 0.000004 \\
\textit{Neisseriales} & 0.029 & 0.013 & 2.1 & 0.000002 & 0.000006 \\
\textit{Turicibacterales} & 0.006 & 0.013 & 0.45 & 0.000008 & 0.00002 \\
\textit{Bifidobacteriales} & 0.041 & 0.09 & 0.47 & 0.00004 & 0.0001 \\
\textit{Bacteroidales} & 25.5 & 38.8 & 0.66 & 0.00008 & 0.00019 \\
\textit{Gemellales} & 0.026 & 0.015 & 1.7 & 0.00023 & 0.00048 \\
\textit{Verrucomicrobiales} & 0.017 & 0.036 & 0.48 & 0.0016 & 0.003 \\
\textit{Sphingomonadales} & 0.010 & 0.007 & 1.4 & 0.02 & 0.04 \\
\textit{Burkholderiales} & 1.3 & 0.86 & 1.6 & 0.02 & 0.04 \\ \\
\multicolumn{6}{c}{\textbf{Family level}} \\
\textit{Lachnospiraceae} & 4.9 & 11.5 & 0.42 & 0 & 0 \\
\textit{Erysipelotrichaceae} & 0.44 & 1.3 & 0.34 & 0 & 0 \\
\textit{Clostridiaceae} & 0.11 & 0.42 & 0.25 & 0 & 0 \\
\textit{Pasteurellaceae} & 1.3 & 0.3 & 4.2 & 0 & 0 \\
\textit{Fusobacteriaceae} & 0.25 & 0.08 & 3.3 & 0 & 0 \\
\textit{Enterobacteriaceae} & 2.8 & 0.84 & 3.4 & 0 & 0.000001 \\
\textit{Neisseriaceae} & 0.029 & 0.014 & 2.1 & 0.000002 & 0.00001 \\
\textit{Ruminococcaceae} & 5.3 & 9.9 & 0.54 & 0.000002 & 0.00001 \\
\textit{Turicibacteraceae} & 0.006 & 0.013 & 0.44 & 0.000006 & 0.00002 \\
\textit{Bifidobacteriaceae} & 0.04 & 0.09 & 0.46 & 0.00003 & 0.0001 \\
\textit{Campylobacteraceae} & 0.013 & 0.007 & 1.7 & 0.00012 & 0.0004 \\
\textit{Christensenellaceae} & 0.007 & 0.01 & 0.55 & 0.00015 & 0.0005 \\
\textit{Porphyromonadaceae} & 0.39 & 0.81 & 0.48 & 0.0002 & 0.0005 \\
\textit{Gemellaceae} & 0.026 & 0.016 & 1.7 & 0.0003 & 0.0009 \\
\textit{Bacteroidaceae} & 21.6 & 32.8 & 0.66 & 0.0004 & 0.001 \\
\textit{Veillonellaceae} & 1.4 & 0.88 & 1.5 & 0.001 & 0.002 \\
\textit{Verrucomicrobiaceae} & 0.018 & 0.038 & 0.47 & 0.001 & 0.003 \\
\textit{Micrococcaceae} & 0.014 & 0.010 & 1.4 & 0.009 & 0.018 \\
\textit{Alcaligenaceae} & 1.0 & 0.58 & 1.7 & 0.02 & 0.03 \\
\textit{Prevotellaceae} & 0.04 & 0.07 & 0.58 & 0.02 & 0.04 \\ \hline
\end{tabular}
\end{table}
\clearpage

\begin{table}[h]
\small
\centering

\begin{tabular}{@{}llllll@{}}
\hline \hline 
\textbf{taxon name} & \textbf{\begin{tabular}[c]{@{}l@{}}abundance, \\ $\bm{l_{\mathrm{CD}}}$ \end{tabular}} & \textbf{\begin{tabular}[c]{@{}l@{}}abundance,\\ $\bm{l_{\mathrm{ctrl}}}$\end{tabular}} & \textbf{\begin{tabular}[c]{@{}l@{}}ratio,\\ $\bm{l_{\mathrm{CD}}/l_{\mathrm{ctrl}}}$\end{tabular}} & \textbf{p-value} & \textbf{q-value} \\ \hline \hline \\

\multicolumn{6}{c}{\textbf{Genus level}} \\ 
\textit{Roseburia} & 0.042 & 0.20 & 0.21 & 0 & 0 \\
\textit{Blautia} & 0.17 & 0.52 & 0.33 & 0 & 0 \\
\textit{Aggregatibacter} & 0.11 & 0.022 & 5.0 & 0 & 0 \\
\textit{Haemophilus} & 1.41 & 0.33 & 4.3 & 0 & 0 \\
\textit{Lachnospira} & 0.022 & 0.076 & 0.29 & 0 & 0 \\
\textit{Actinobacillus} & 0.025 & 0.009 & 2.7 & 0 & 0 \\
\textit{Fusobacterium} & 0.36 & 0.10 & 3.7 & 0 & 0 \\
\textit{Coprococcus} & 0.35 & 0.87 & 0.40 & 0 & 0 \\
\textit{[Eubacterium]} & 0.048 & 0.13 & 0.36 & 0 & 0 \\
\textit{Veillonella} & 0.30 & 0.13 & 2.2 & 0.000001 & 0.000006 \\
\textit{Campylobacter} & 0.018 & 0.009 & 1.9 & 0.000002 & 0.000009 \\
\textit{Eikenella} & 0.018 & 0.009 & 2.1 & 0.000002 & 0.000009 \\
\textit{Neisseria} & 0.019 & 0.010 & 1.9 & 0.000002 & 0.000009 \\
\textit{Faecalibacterium} & 1.92 & 4.27 & 0.45 & 0.000003 & 0.000009 \\
\textit{Erwinia} & 0.016 & 0.009 & 1.9 & 0.000024 & 0.000076 \\
\textit{Dialister} & 0.25 & 0.091 & 2.7 & 0.000035 & 0.0001 \\
\textit{Holdemania} & 0.02 & 0.036 & 0.54 & 0.000039 & 0.0001 \\
\textit{Turicibacter} & 0.008 & 0.017 & 0.5 & 0.00015 & 0.0004 \\
\textit{[Ruminococcus]} & 0.57 & 0.91 & 0.62 & 0.00018 & 0.0004 \\
\textit{Ruminococcus} & 0.57 & 0.91 & 0.62 & 0.00018 & 0.0004 \\
\textit{Parabacteroides} & 0.44 & 0.91 & 0.49 & 0.0003 & 0.0008 \\
\textit{Bifidobacterium} & 0.058 & 0.11 & 0.53 & 0.0007 & 0.001 \\
\textit{Rothia} & 0.016 & 0.011 & 1.5 & 0.0008 & 0.002 \\
\textit{Porphyromonas} & 0.018 & 0.010 & 1.7 & 0.001 & 0.002 \\
\textit{Sutterella} & 1.46 & 0.73 & 2.0 & 0.002 & 0.004 \\
\textit{Dorea} & 0.48 & 0.73 & 0.66 & 0.002 & 0.004 \\
\textit{Bacteroides} & 1.22 & 41.9 & 0.75 & 0.005 & 0.01 \\
\textit{Akkermansia} & 0.023 & 0.044 & 0.53 & 0.006 & 0.01 \\
\textit{Anaerostipes} & 0.012 & 0.018 & 0.7 & 0.01 & 0.02 \\
\textit{Staphylococcus} & 0.02 & 0.014 & 1.4 & 0.02 & 0.03 \\
\textit{Granulicatella} & 0.034 & 0.024 & 1.4 & 0.02 & 0.03 \\
\textit{Phascolarctobacterium} & 0.038 & 0.061 & 0.62 & 0.03 & 0.04 \\ \\
\multicolumn{6}{c}{\textbf{Species level}} \\
H. parainfluenzae & 3.42 & 0.83 & 4.1 & 0 & 0 \\
\textit{A. segnis} & 0.064 & 0.023 & 2.8 & 0 & 0 \\
\textit{F. prausnitzii} & 5.0 & 12.3 & 0.41 & 0 & 0.000003 \\
\textit{B. adolescentis} & 0.028 & 0.066 & 0.43 & 0.000005 & 0.00004 \\
\textit{E. dolichum} & 0.10 & 0.23 & 0.44 & 0.000007 & 0.00004 \\
\textit{V. parvula} & 0.06 & 0.033 & 1.82 & 0.00002 & 0.0001 \\
\textit{V. dispar} & 0.51 & 0.27 & 1.91 & 0.0002 & 0.0008 \\
\textit{N. subflava} & 0.041 & 0.025 & 1.62 & 0.0008 & 0.0027 \\
\textit{Ros. faecis} & 0.023 & 0.035 & 0.65 & 0.0008 & 0.0027 \\
\textit{P. copri} & 0.052 & 0.11 & 0.46 & 0.001 & 0.003 \\
\textit{A. muciniphila} & 0.061 & 0.13 & 0.48 & 0.002 & 0.006 \\
\textit{Bac. uniformis} & 0.71 & 1.2 & 0.58 & 0.012 & 0.027 \\
\textit{R. mucilaginosa} & 0.039 & 0.028 & 1.39 & 0.015 & 0.031 \\
\textit{Bl. producta} & 0.031 & 0.046 & 0.67 & 0.015 & 0.031 \\
\textit{C. catus} & 0.045 & 0.067 & 0.67 & 0.021 & 0.039 \\ \hline
\end{tabular}
\end{table}

\clearpage
\begin{table}[]
\small
\centering
\caption{\textbf{A summary of interaction strengths and log-abundance correlation coefficients for the core IBD network shown in Fig.~3 of the main text}. Statistical significance was estimated by a permutation test. Specifically, we independently permuted the abundance of each taxa across samples and then computed the correlation and interaction matrices on the permuted data to generate the probability distribution for the null hypothesis of no interaction.}
\label{tab:interactions}
\begin{tabular}{@{}lllll@{}}
\hline \hline
\textbf{interacting taxa} & \textbf{\begin{tabular}[c]{@{}l@{}}correlation\\ strength, $C_{ij}$\end{tabular}} & \textbf{\begin{tabular}[c]{@{}l@{}}interaction \\ strength, $J_{ij}$\end{tabular}} & \textbf{\begin{tabular}[c]{@{}l@{}}q-value, \\ correlation\end{tabular}} & \textbf{\begin{tabular}[c]{@{}l@{}}q-value, \\ interaction\end{tabular}} \\ \hline \hline
 \\
\textit{A.segnis-B.producta} & +0.16 & +0.14 & 0.0011 & 0.0041 \\
\textit{A.segnis-Oscillospira} & $-$0.16 & $-$0.17 & 0.0014 & 0.0011 \\
\textit{A.segnis-Roseburia} & $-$0.15 & $-$0.19 & 0.0034 & 0.0006 \\
\textit{A.segnis-Sutterella} & $-$0.015 & +0.046 & 0.80 & 0.41 \\
\textit{A.segnis-Turicibacter} & +0.18 & +0.12 & 0 & 0.021 \\
\textit{B.adolescentis-A.segnis} & +0.19 & +0.19 & 0 & 0.0006 \\
\textit{B.adolescentis-B.producta} & +0.26 & +0.16 & 0 & 0.0019 \\
\textit{B.adolescentis-Oscillospira} & +0.069 & $-$0.067 & 0.17 & 0.24 \\
\textit{B.adolescentis-Roseburia} & +0.25 & +0.24 & 0 & 0 \\
\textit{B.adolescentis-Sutterella} & +0.036 & +0.055 & 0.50 & 0.34 \\
\textit{B.adolescentis-Turicibacter} & +0.40 & +0.46 & 0 & 0 \\
\textit{B.producta-Oscillospira} & +0.10 & +0.04 & 0.044 & 0.47 \\
\textit{B.producta-Roseburia} & +0.100 & +0.0063 & 0.047 & 0.92 \\
\textit{B.producta-Sutterella} & +0.0012 & +0.092 & 0.98 & 0.091 \\
\textit{B.producta-Turicibacter} & +0.31 & +0.23 & 0 & 0 \\
\textit{E.dolichum-A.segnis} & $-$0.0063 & $-$0.027 & 0.92 & 0.66 \\
\textit{E.dolichum-B.adolescentis} & +0.19 & +0.051 & 0.0002 & 0.35 \\
\textit{E.dolichum-B.producta} & +0.40 & +0.46 & 0 & 0 \\
\textit{E.dolichum-F.prausnitzii} & +0.075 & +0.0087 & 0.13 & 0.92 \\
\textit{E.dolichum-Oscillospira} & +0.27 & +0.29 & 0 & 0 \\
\textit{E.dolichum-Roseburia} & +0.25 & +0.21 & 0 & 0 \\
\textit{E.dolichum-Sutterella} & $-$0.080 & $-$0.19 & 0.11 & 0 \\
\textit{E.dolichum-Turicibacter} & +0.20 & +0.057 & 0 & 0.33 \\
\textit{F.prausnitzii-A.segnis} & $-$0.086 & +0.0064 & 0.086 & 0.92 \\
\textit{F.prausnitzii-B.adolescentis} & +0.15 & +0.20 & 0.0021 & 0 \\
\textit{F.prausnitzii-B.producta} & $-$0.065 & $-$0.15 & 0.19 & 0.0032 \\
\textit{F.prausnitzii-Oscillospira} & +0.32 & +0.29 & 0 & 0 \\
\textit{F.prausnitzii-Roseburia} & +0.35 & +0.35 & 0 & 0 \\
\textit{F.prausnitzii-Sutterella} & +0.25 & +0.204 & 0 & 0.0006 \\
\textit{F.prausnitzii-Turicibacter} & $-$0.095 & $-$0.18 & 0.053 & 0.0003 \\
\textit{Roseburia-Oscillospira} & +0.29 & +0.16 & 0 & 0.0034 \\
\textit{Roseburia-Sutterella} & +0.099 & +0.019 & 0.05 & 0.76 \\
\textit{Roseburia-Turicibacter} & +0.099 & +0.053 & 0.05 & 0.34 \\
\textit{Sutterella-Oscillospira} & +0.23 & +0.24 & 0 & 0 \\
\textit{Turicibacter-Oscillospira} & +0.036 & +0.076 & 0.50 & 0.18 \\
\textit{Turicibacter-Sutterella} & $-$0.12 & $-$0.15 & 0.012 & 0.0026 \\\\ \hline\hline
\end{tabular}
\end{table}

\clearpage

\end{document}